\documentclass[fleqn,usenatbib]{mnras}

\usepackage{newtxtext,newtxmath}
\usepackage{diagbox}
\usepackage{stfloats}
\usepackage{lineno}
\usepackage{color}

\usepackage[T1]{fontenc}
\usepackage{ae,aecompl}
\usepackage{xspace}
\usepackage{CJK}

\usepackage{graphicx}	
\usepackage{subfigure}
\usepackage{float}
\usepackage{amsmath}	
\usepackage{amssymb}	
\usepackage{ltxtable}
\usepackage[normalem]{ulem}

\newcommand{\be}{\begin{equation}}
\newcommand{\ee}{\end{equation}}
\newcommand{\ba}{\begin{eqnarray}}
\newcommand{\ea}{\end{eqnarray}}

\newcounter{ichi}
\newcounter{ni}
\newcounter{san}
\newcounter{yon}
\title[Multi-wavelength Afterglows of FRB and magnetars]{Multi-wavelength afterglow emission from bursts associated with magnetar flares and fast radio bursts}

\author[Wei et al.]{
Yujia Wei (魏煜佳)$^{1,2,3}$, 
B. Theodore Zhang (张兵)$^{4}$,
Kohta Murase$^{5,6,7,8,4}$
\\
$^1$Key Laboratory of Dark Matter and Space Astronomy, Purple Mountain Observatory, Chinese Academy of Sciences, Nanjing 210023, China\\
$^2$School of Astronomy and Space Science, University of Science and Technology of China, Hefei 230026, China\\
$^3$School of Astronomy \& Space Science, Nanjing University, Nanjing 210023, China\\
$^4$Center for Gravitational Physics and Quantum Information, Yukawa Institute for Theoretical Physics, Kyoto University, Kyoto 606-8502, Japan\\
$^5${Department of Physics, The Pennsylvania State University, University Park, PA 16802, USA} \\
$^6${Department of Astronomy \& Astrophysics, The Pennsylvania State University, University Park, PA 16802, USA} \\
$^7${Center for Multimessenger Astrophysics, Institute for Gravitation and the Cosmos, The Pennsylvania State University, University Park, PA 16802, USA} \\
$^8$School of Natural Sciences, Institute for Advanced Study, Princeton, NJ 08540, USA \\
}

\date{Accepted XXX. Received YYY; in original form ZZZ}

\pubyear{2022}

\begin{document}
\begin{CJK*}{UTF8}{gbsn}

\label{firstpage}
\pagerange{\pageref{firstpage}--\pageref{lastpage}}
\maketitle

\begin{abstract}
Magnetars have been considered as progenitors of magnetar giant flares (MGFs) and fast radio bursts (FRBs). We present detailed studies on afterglow emissions caused by bursts that occur in their wind nebulae and surrounding baryonic ejecta. In particular, following the bursts-in-bubble model proposed by Murase, Kashiyama \& M\'esz\'aros, we analytically and numerically calculate spectra and light curves of such afterglow emission. We scan parameter space for the detectability of radio signals, and find that a burst with $\sim10^{45}~{\rm erg}$ is detectable with the Very Large Array or other next-generation radio facilities. 
The detection of multi-wavelength afterglow emission from MGFs and/or FRBs is of great significance for their localization and revealing their progenitors, and we estimate the number of detectable afterglow events. 
\end{abstract}




\section{Introduction}
Magnetars are young, highly magnetized neutron stars~\citep[e.g.,][for a review]{Kaspi_2017ARA&A..55..261K}. Their emission is typically observed at X-ray and soft gamma-ray bands, which are powered by the decay of magnetic fields. The magnetars can be phenomenologically divided into two classes, which are anomalous X-ray pulsars and soft gamma repeaters (SGRs), and SGRs are characterized by weak and recurring short bursts. They are known to produce giant flares in the gamma-ray band, which we call magnetar giant flares (MGFs). A small fraction of gamma-ray bursts (GRBs) originate from MGFs, and examples include GRB 790305B~\citep[e.g.,][]{Mazets_1979Natur.282..587M}, GRB 980827~\citep[e.g.,][]{Mazets_1999AstL...25..635M, Hurley_1999Natur.397...41H}, and GRB 041227~\citep[e.g.,][]{Palmer_2005Natur.434.1107P, Frederiks_2007AstL...33....1F} for Galactic events, and GRB 200415A has been thought to be a most likely extragalactic MGF~\citep[e.g.,][]{Yang_2020ApJ...899..106Y,Minaev_2020AstL...46..573M,Roberts_2021Natur.589..207R,Svinkin_2021Natur.589..211S, Zhang:2022npz}.

Magnetars have been thought to be the most promising progenitors of fast radio bursts (FRBs)~\citep[e.g.,][for reviews]{Petroff_2019A&ARv..27....4P, Zhang_2022arXiv221203972Z}. In 2020, FRB 200428 detected by CHIME~\citep{2020Nature..587..54} and STARE2~\citep{Bochenek_2020} was confirmed to be associated with a Galactic magnetar SGR 1935+2154. This evidence indicates that magnetars are at least one of the origins of FRBs. 

Detecting multi-wavelength counterparts of FRBs is crucial for better understanding the progenitors and mechanisms, and more dedicated multi-wavelength search campaigns are necessary~\citep{Nicastro_2021Univ....7...76N}. 
However, theoretical predictions are highly model dependent. Among various possibilities, possible connections between FRBs and gamma-ray transients have often been discussed~\citep[e.g.,][]{Popov&Postnov10, Zhang_2016ApJ...822L..14Z, Murase_2017ApJ...836L...6M}, but clear associations have not been established so far~\citep[e.g.,][]{DeLaunay+16frb131104, MAGIC:2018bjc, Martone:2019aeo, Venere:2021hdv}. 
Detecting longer-lasting afterglow emission from outflows launched with transient activities is also helpful for identifying the FRB counterparts. Some studies have been proposed to predict the detectability of afterglow emission based on GRB-like models. For example, \cite{Yi_2014_ApJ} studied multi-wavelength afterglows from GRB-like outflows, and \cite{Lin_2020MNRAS.498.2384L} studied the detectability of afterglows in light of the binary neutron star merger model for FRBs. \cite{Li_2022ApJ...929..139L} discussed a possible connection between FRB 180916B and a possible optical counterpart AT2020hur of FRB assuming a collimated outflow.  

In this work, following the bursts-in-bubble model proposed by~\cite{Murase:2016sqo}, we study week-to-month scale afterglow emission caused by outflows that may be associated with MGFs and/or FRBs. 
Magnetars may inject rotation and/or magnetic energy into their nebulae. If there is a burst (e.g., FRB and/or MGF) inside such a nebula, an associated outflow will sweep the nebula with a mass of $\sim10^{-9}-10^{-5}~M_\odot$~\citep{Murase:2016sqo}, forming trans-relativistic ejecta. This may be consistent with radio afterglow emission observed from SGR 1806-20, for which a fading radio source VLA J180839-202439 was identified \citep{gs06}.  
According to~\cite{Kaspi_2017ARA&A..55..261K}, of the 23 magnetars identified, eight are reliably related to supernova (SN) remnants. If magnetars are as young as $
\sim10-100$~yr, it is rather natural to expect that the outflow interacts with dense baryonic SN ejecta. 

We use both analytical and numerical methods to calculate the dynamic evolution of the refreshed forward shock and the corresponding afterglow emission.
We also study whether such afterglow emission from a nearby galaxy like M 81 can be detected in the radio, optical and X-ray bands. We then scan the parameter space to find the detection horizon for current and next-generation radio telescopes, which include the Atacama Large Millimeter/sub-millimeter Array (ALMA), the Very Large Array (VLA), the Square Kilometer Array (SKA) and the Next Generation Very Large Array (ngVLA). Finally, We calculate the expected number of afterglow events detected with these radio telescopes.

We use notations as $Q_x=Q/10^x$ in the CGS unit except $t_{\rm yr}\equiv (t/1~{\rm yr})$ and $M_{\rm ej,1 M_\odot}=M_{\rm ej}/1 M_\odot$.

\section{Model}
\label{sect:method}
In this section, we describe the details of the model and methods that are used to calculate multi-wavelength afterglow emission.

\begin{figure}
	\centering
	\subfigure[]{
	    \begin{minipage}[b]{0.4\textwidth}
	    \begin{center}
	    \includegraphics[width=1.0\textwidth]{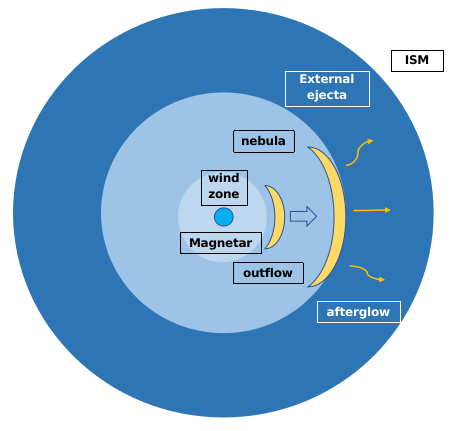}
	    \end{center}
	    \end{minipage}	    
	}%
	\caption{A schematic picture of afterglow emission following a burst, which may be associated with an MGF and/or FRB. A magnetar is surrounded by its wind nebula and baryonic SN/merger ejecta (external baryonic ejecta). A burst-driven outflow sweeps the nebula and forms a trans-relativistic shell, which is quickly decelerated by the dense baryonic ejecta, and afterglow emission is generated by the refreshed forward shock.}
\label{fig:model}
\end{figure}

In Fig.~\ref{fig:model}, we show the schematic picture of the bursts-in-bubble model~\citep{Murase:2016sqo}. 
The magnetic and/or spin-down activity of magnetars powers relativistic winds that are accelerated mainly in the wind zone between the light cylinder and the nebula, leading to the formation of a wind bubble embedded in the SN or merger ejecta. 
Then an impulsive burst that may be associated with an MGF and/or FRB may be accompanied by an outflow that will sweep up the nebula. The pre-existing non-thermal particles in the nebula may be boosted by the outflow, which may lead to the gamma-ray emission~\citep[e.g.,][]{Lyubarsky14, Murase:2016sqo}. 
The swept-up nebula can still be trans-relativistic, which may eventually interact with the baryonic SN/merger ejecta, by which afterglow emission is expected, as shown in Fig.~\ref{fig:model}. 

\subsection{Dynamics}

At early times, the merged shell is in the free-expansion phase where the velocity $v_s$ of the forward shock keeps constant and its value is equal to the initial velocity of the merged shell. During this free-expansion phase, the radius of the shock is
\begin{equation}
\label{eq:ana_r_free}
r_s \approx v_0 t \simeq 2.6 \times 10^{15} \ {\rm cm} \ t_{5}, \ \ \ (t \leq t_{\rm dec}),
\end{equation}
where $t$ is the observation time after the burst and $v_0 \simeq 2.6 \times 10^{10} \rm~cm~s^{-1}$ is the initial velocity of the shock corresponding to the initial Lorentz factor $\Gamma_0 = 2$. 
When the accumulated mass is equal to the initial mass of the merged shell, the merged shell begins to decelerate. This deceleration time $t_{\rm dec}$ and the corresponding deceleration radius $r_{\rm dec}$ are~\citep[e.g.,][]{2011Natur.478...82N}
\begin{equation}\label{eq:dec}
	\begin{aligned}
		& r_{\rm dec} = {\left(\frac{3 \mathcal{E}_{k}}{4 \pi n_{\rm ext} m_p c^2 \Gamma_0^2}\right)}^{1/3} \simeq 1.6 \times 10^{15} \ {\rm cm} \ \mathcal{E}_{k,47}^{1/3} n_{\rm ext,3}^{-1/3},\\
		& t_{\rm dec} = \frac{r_{\rm dec}}{v_0} \simeq 6.1 \times 10^{4} {\rm~s} \ \mathcal{E}_{k,47}^{1/3} n_{\rm ext,3}^{-1/3},
	\end{aligned}
\end{equation}
where $\mathcal{E}_k$ is the kinetic energy of the merged shell, $n_{\rm ext}$ is the number density of the external baryonic ejecta.
This number density can be estimated through the following formula
\begin{equation}
\label{eq:n_ext}
	\begin{aligned}
	n_{\rm ext} & = \frac{3 M_{\rm ext}}{4 \pi R_{\rm ext}^3 m_p}\\
    & \simeq 2.9 \times 10^5 \ {\rm cm^{-3}} \ M_{\rm ext, 1 M_\odot} V_{\rm ext,8.5}^{-3} T_{\rm age, \rm 10~yr}^{-3}, 
	\end{aligned}
\end{equation}
where $R_{\rm ext} = V_{\rm ext} T_{\rm age} \simeq 1.0 \times 10^{17} \ {\rm cm} \ V_{\rm ext,8.5} T_{\rm age, 10 {\rm~yr}}$ is the radius of SN ejecta, $V_{\rm ext}$ is the velocity of SN ejecta, $M_{\rm ext}$ is the mass of SN ejecta, and $T_{\rm age}$ is the age of the magnetar at the time of a burst.  

During the deceleration phase, the evolution of the forward shock following the Sedov-Taylor solution, where the shock velocity is 
\begin{equation}\label{eq:vs}
	v_s \simeq 3.7 \times 10^9 {\rm cm \ s^{-1}} \ \mathcal{E}^{1/5}_{k,47} \, n_{\rm ext, 3}^{-1/5} t_{6}^{-3/5},\ \ \ (t > t_{\rm dec}),
\end{equation}
and the shock radius $r_s$ is 
\begin{equation}
		 r_s \simeq 1.0 \times 10^{16} \ {\rm cm} \ \mathcal{E}_{k,47}^{1/5} \, n_{\rm ext,3}^{-1/5} t_{6}^{2/5},\ \ \ (t > t_{\rm dec}).
\end{equation}
Note the above Sedov-Taylor solution will not be valid when the radius of the forward shock is larger than the extension of the external baryonic ejecta, which means $r_s$ should be smaller than $R_{\rm ext}$.

The above estimates are not accurate in the trans-relativistic case and become invalid in the relativistic case, where the shock propagates a distance of $dr = \beta c / (1 - \beta) dt$ during the time interval $dt$ when measured in the observer frame. In order to have better modeling of the transition from the free-expansion phase to the deceleration phase, we numerically solve the dynamical evolution of the interacting shell~\citep[e.g.,][]{Nava_2013MNRAS.433.2107N}.
The total energy of the interacting shell is
\begin{equation}
\mathcal{E}_{\rm tot} = \Gamma M_{\rm ej} c^2 + \Gamma m_{\rm ext} c^2 + \frac{\hat{\gamma}\Gamma^2 - \hat{\gamma} + 1}{\Gamma} \mathcal{E}_{\rm int}^\prime,
\end{equation}
where $\mathcal{E}_{\rm tot}$ is the total energy, $\Gamma$ is the Lorentz factor of the merged shell, $M_{\rm ej}$ is the mass of the merged shell, $m_{\rm ext}$ is the mass of accumulated external baryonic ejecta, $\hat{\gamma} = (4 + \Gamma^{-1})/3$ is the adiabatic index, and $\mathcal{E}_{\rm int}^\prime$ is the internal energy. Considering energy conservation, i.e., $d\mathcal{E}_{\rm tot} = dm_{\rm ext} c^2$ where the change of $\mathcal{E}_{\rm tot}$ is due to the accumulation of the external baryonic ejecta, we can establish the differential equation to describe the evolution of $\Gamma$. See the details in~\cite{zhang_external_2021}.

\begin{figure*}
	\subfigure[]{
		\includegraphics[scale=0.6]{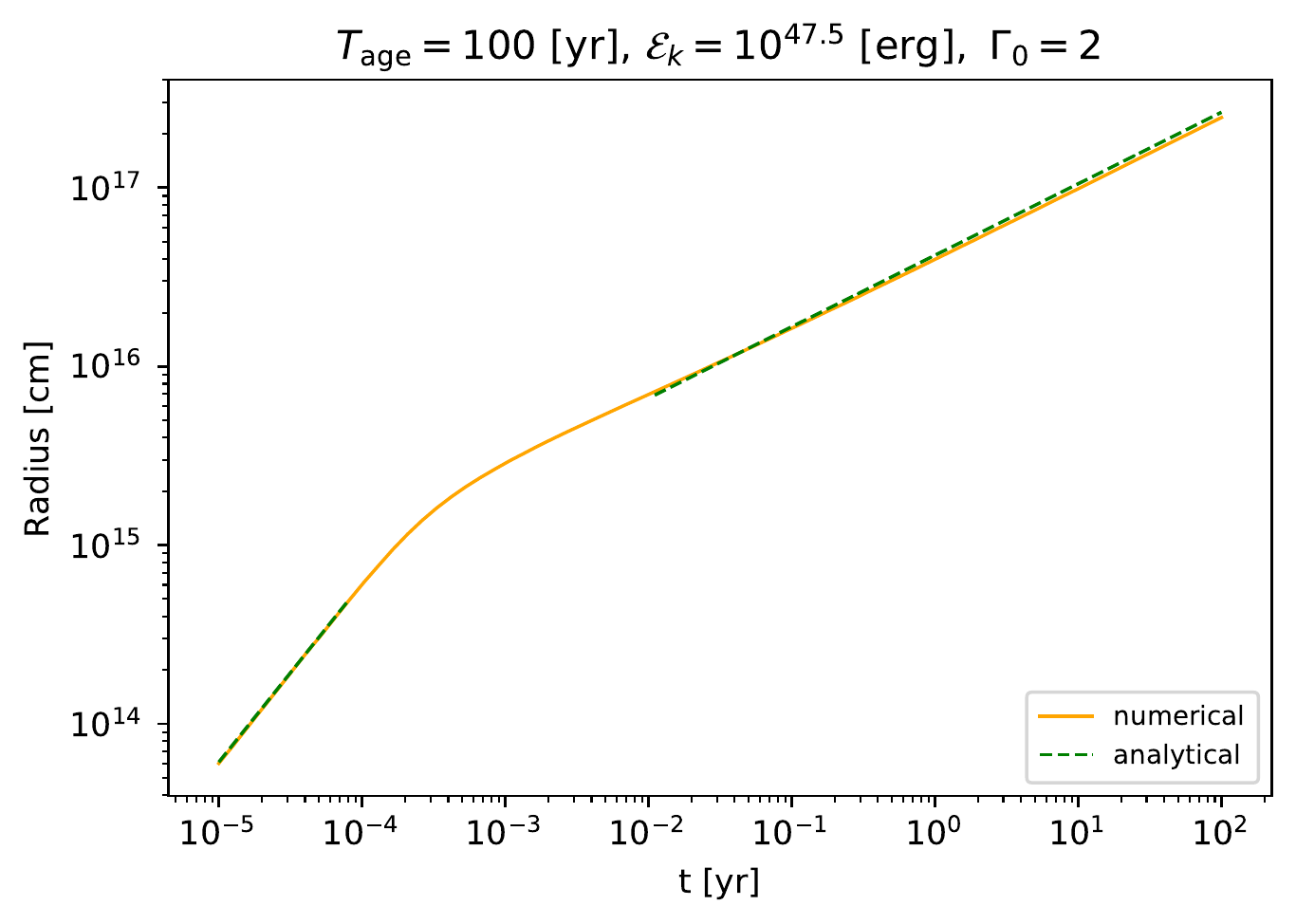}
	}%
	\subfigure[]{
		\includegraphics[scale=0.6]{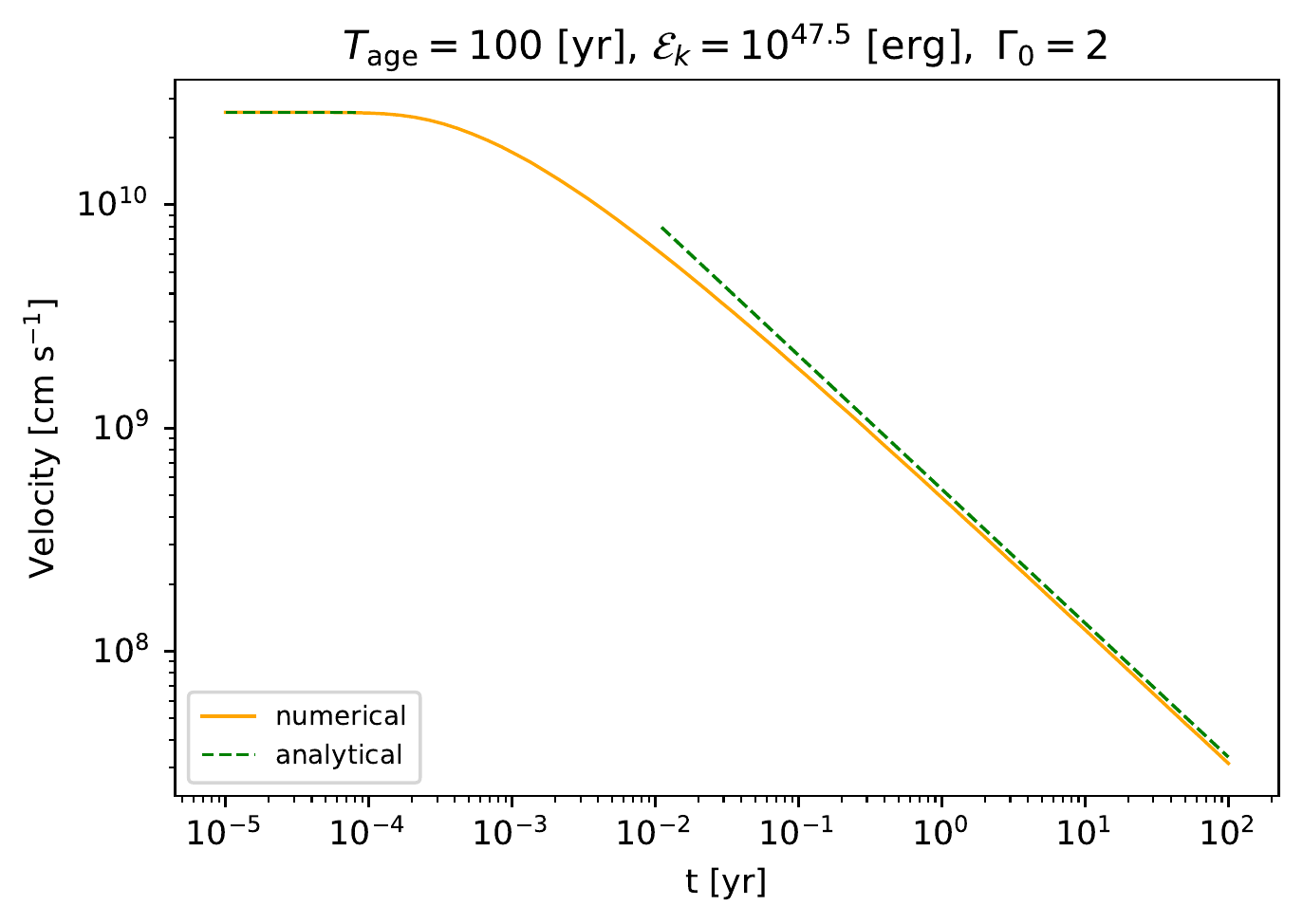}
	}%
	\caption{Analytical and numerical results for the thin-shell approximation on the dynamic evolution of the forward shock caused by interaction with the external baryonic ejecta. The yellow line represents the numerical result, while the green line indicates the analytical result. 
 }
	\label{fig:ananum_dynamic}
\end{figure*}

Throughout the work we assume $\Gamma_0 = 2$, corresponding to $v_0 \simeq 0.87 c$. We compare the analytical (green lines) and numerical (yellow lines) dynamic evolution for $\mathcal{E}_k = 10^{47.5} {\rm~erg}$ and $T_{\rm age} = 100 {\rm~yr}$ in Fig.~\ref{fig:ananum_dynamic}. For analytical curves, we do not show the result in the transition phase since it is not accurate. However, the free expansion phase and the deceleration phase could be smoothly connected via solving the dynamical evolution of the interacting shell numerically. 

In above case, we adopt the thin-shell approximation of the inner ejecta, which are formed when energetic relativistic outflow merged into the nebula. According to the burst-in-bubble model \citep{Murase:2016sqo}, after the burst, the shocked nebula is highly compressed and confined into a thin layer with a width depending on the ratio of the pressure between the energetic relativistic outflow and the nebula. To take into account the thick shell evolution, we modify the differential equation on the evolution of the Lorentz factor Eq. A2 used in ~\cite{zhang_external_2021}. In particular, we use the method proposed in~\cite{Nava_2013MNRAS.433.2107N}, and assume that the ejecta have a width of $\Delta R_{\rm ej}=R_{\rm nb}/ 12\Gamma_0^2$ measured in the engine frame and $t_{\rm cross} = \Delta R_{\rm ej} / v_0$ is the crossing time of the reverse shock without spreading. The emission will peak at $t_{\times} = {\rm max} [t_{\rm cross}, t_{\rm dec}]$.

\subsection{Synchrotron emission}
Magnetic fields may be amplified via the shock, as indicated in GRBs~\citep[e.g.,][]{Santana_2014ApJ...785...29S, Mizuno_2014MNRAS.439.3490M}. The downstream magnetic field is estimated to be
\begin{equation}
	\begin{aligned}
		B_{\rm fs} & \approx (9 \pi \epsilon_B n_{\rm ext} m_p {v_s}^2)^{1/2}\\
		& \simeq 0.026 \ {\rm G} \ \mathcal{E}_{k,47}^{1/5} n_{\rm ext,3}^{3/10}  t_{6}^{-3/5} \epsilon_{B,-3}^{1/2},
	\end{aligned}
\end{equation}
where $\epsilon_B$ is the magnetic energy fraction transferred from post-shock thermal energy. 
The minimum Lorentz factor of the accelerated electron is 
\begin{equation}\label{eq:gamma_min}
	\begin{aligned}
	\gamma_{m} - 1 & \approx {\rm max}~\left[1, \frac{m_p \epsilon_{e} (s-2)}{2 m_e f_{e} (s-1)} \left(\frac{v_s}{c}\right)^2\right] \\
		&\simeq {\rm max}~\left[1, 0.24 \ \mathcal{E}_{k,47}^{2/5} n_{\rm ext,3}^{-2/5}  t_{6}^{-6/5} \epsilon_{e,-1} f_{e,0}^{-1}\right],
		\end{aligned}
\end{equation}
where $s = 2.2$ is the spectral index, $\epsilon_{e}$ is the fraction of post-shock thermal energy transferred to electron energy, and $f_{e}$ is the fraction of thermal electrons accelerated to the non-thermal distribution. Note that the minimum Lorentz factor is fixed to $\gamma_{m, \rm min} = 2$ in the deep-Newtonian phase~\citep[e.g.,][]{huang_gamma-ray_2003, granot_diagnosing_2006, sironi_late-time_2013}.
The corresponding minimum synchrotron frequency is
\begin{equation}
	\begin{aligned}
		\nu_{m} & = \frac{3}{4\pi} \gamma_{m}^2 \frac{e B_{\rm fs}}{m_{e} c} \\
		& \simeq 1.6 \times 10^5 \ {\rm Hz} \ \mathcal{E}_{k,47} n_{\rm ext,3}^{-1/2}  t_{6}^{-3} \epsilon_{ B,-3}^{1/2} \epsilon_{ e,-1}^2 f_{e,0}^{-2}, 
	\end{aligned}
\end{equation}
where $e$ is the electron charge. Note we use $\gamma_m = 1 + \frac{m_p \epsilon_{e} (s-2)}{2 m_e f_{e} (s-1)} \left(\frac{v_s}{c}\right)^2$ to derive the second line of above equation but here $\gamma_m$ should be $2$.
The synchrotron cooling frequency is
\begin{equation}
	\begin{aligned}
		\nu_{c} & = \frac{3}{4\pi} \gamma_{c}^2 \frac{e B_{\rm fs}}{m_e c} \\
		& \simeq 1.5 \times 10^{17} \ {\rm Hz}\ \mathcal{E}_{k,47}^{-3/5} n_{\rm ext,3}^{-9/10} t_{6}^{-1/5} \epsilon_{ B,-3}^{-3/2},
	\end{aligned}
\end{equation}
where $\gamma_c = 6 \pi m_{e} c / (\sigma_{T} B_{\rm fs}^2 t)$ is the synchrotron cooling Lorentz factor.
The maximum synchrotron frequency is
\begin{equation}\label{eq:nu_max}
	\begin{aligned}
		\nu_{\rm max} & = \frac{3}{4\pi} \gamma_{\rm max}^2 \frac{e B_{\rm fs}}{m_{e} c} \\
		& \simeq 1.3 \times 10^{20} \ {\rm Hz}\ \mathcal{E}_{k,47}^{2/5} n_{\rm ext,3}^{-2/5} t_{6}^{-6/5},
	\end{aligned}
\end{equation}
where $\gamma_{\rm max} = \sqrt{9 \pi v_s^2 e / (10 \sigma_{T} B_{\rm fs} c^2)}$ is the maximum Lorentz factor.

The synchrotron peak flux $F_{\nu}^{\rm max}$ is
\begin{equation}\label{eq:ana_F_nu^max}
F_{\nu}^{\rm max} \simeq 0.6 \frac{\sqrt{3} e^3 B_{\rm fs}}{m_{e} c^2} \frac{4 \pi r_s^3 f_e n_{\rm ext} }{3 \times 4 \pi d_{L}^2}
\begin{cases}
1 & \beta \gg \beta_{\rm DN} \\
{\left(\frac{\beta^2}{\beta_{\rm DN}^2}\right)}^{\frac{s-1}{2}}& \beta \ll \beta_{\rm DN}
\end{cases}
\end{equation}
where $d_{L}$ is the luminosity distance from the observer to the source and $\beta_{\rm DN} = \sqrt{(2 m_e f_e (s-1))/(m_p \epsilon_e (s-2))} \simeq 0.26 f_{e,0}^{1/2} \epsilon_{e,-1}^{-1/2}$ is the critical velocity below which the interacting shell is in the deep-Newtonian phase~\citep[e.g.,][]{sironi_late-time_2013, matsumoto_radio_2021}. Note that when $t = t_{\rm DN} \simeq 3.0 \times 10^5 {~\rm s} \ \mathcal{E}_{k, 47}^{1/3} n_{\rm ext, 3}^{-1/3} f_{e,0}^{-5/6} \epsilon_{e,-1}^{5/6}$, the interacting shell will enter into the deep Newtonian regime.

Besides, synchrotron-self absorption (SSA) is important for the attenuation of low-frequency synchrotron emission.
The SSA optical depth can be estimated by $\tau_{\rm ssa} (\nu) = \int \alpha_{\nu} ds$, where $\alpha_{\nu}$ is the SSA coefficient and the integral is along the width of the shocked region.
We can then get the SSA frequency $\nu_{a}$ by using $\tau_{\rm ssa} (\nu_{a}) = 1$,
and we have
\begin{equation}
\begin{aligned}
\nu_{a} 
    & \simeq 7.3 \times 10^8 {\rm~Hz} \ \mathcal{E}_{k,47}^{1/5} n_{\rm ext,3}^{(3s+22)/[10(s+4)]} t_{6}^{(-3s-2)/[5(s+4)]} \\ & \times \epsilon_{B,-3}^{(s+2)/[2(s+4)]} f_{e, 0}^{2 / (s + 4)},
    \end{aligned}
\end{equation} 
for $s = 2.2$ and $\gamma_m = 2$ in the deep Newtonian phase.
Note that the number of electrons contributing to the SSA process in the deep-Newtonian phase should be corrected accordingly similar to Eq.~\ref{eq:ana_F_nu^max}.
The synchrotron emission spectrum can be analytically calculated as
\begin{equation}
\label{eq:flux2}
	F_{\rm \nu} = F_{\nu}^{\rm max} 
	\begin{cases}
		{\left(\frac{\nu}{\nu_{m}}\right)}^2 {\left(\frac{\nu_{a}}{\nu_{m}}\right)}^{-s/2-2} & \nu \leq \nu_{m}\\
		{\left(\frac{\nu}{\nu_{a}}\right)}^{5/2} {\left( \frac{\nu_{a}}{\nu_{m}}\right)}^{-(s-1)/2} & \nu_{m} < \nu \leq \nu_{a} \\
		{\left(\frac{\nu}{\nu_{m}}\right)}^{-(s-1)/2} & \nu_{a} < \nu \leq \nu_{c} \\	
		{\left(\frac{\nu}{\nu_{c}}\right)}^{-s/2} {\left(\frac{\nu_{c}}{\nu_{m}}\right)}^{-(s-1)/2} & \nu_{c} < \nu < \nu_{\rm max}
	\end{cases},
\end{equation}
in the slow cooling case.

The free-free (FF) absorption is also important when the density of the external baryonic ejecta are so high that free electrons jump to higher-energy states via absorbing low-energy photons.
The corresponding optical depth is~\citep[e.g.,][]{Lang_1999acfp.book.....L, Murase_2017ApJ...836L...6M}
\begin{equation}
    \begin{aligned}
            \tau_{\rm ff}(\nu) & \simeq 8.5 \times 10^{-28} \ \bar{Z}^2 (R_{\rm ext} - r_s) {\left(\frac{\nu}{10^{10} \ \rm Hz}\right)}^{-2.1} n_e n_i \\
            & \times {\left(\frac{T_{\rm ext}}{10^4 \ \rm K}\right)}^{-1.35} {\left(\frac{1-e^{-h\nu / kT_{\rm ext}}}{h \nu / k T_{\rm ext}}\right)},
    \end{aligned}
\end{equation}
where $\bar{Z} = 1$ is the average charge number, $T_{\rm ext}$ is the temperature of the external baryonic ejecta, $R_{\rm ext}$ is the size of the external baryonic ejecta, $n_e$ is the free electron number density, and $n_i$ is the ion number density.
In the singly ionized state, we have $n_e = n_{\rm ext} / \mu_e$, and we assume $\mu_e = 1$ for simplicity. Then we take into account the FF absorption by multiplying $\rm{exp}(-\tau_{\rm ff})$. 

\subsection{Numerical calculations}
In the following, we describe the details of our numerical calculations.
We employ the numerical code developed initially for GRB afterglows~\citep{murase_implications_2011,amt20,zhang_external_2021}, but modified it to treat the non-relativistic regime self-consistently\footnote{The code will be made public as a part of the GRB code used in \cite{amt20,zhang_external_2021}.}. 
We obtain the steady-state electron distribution by solving the kinetic equation in the momentum space
\begin{equation}\label{eq:powerlaw}
    \frac{\partial}{\partial p_e}\left(n_{p_e}(t)\frac{dp_e}{dt}\right) + \frac{n_{p_e}(t)}{t_{\rm esc}}  = \dot{n}_{p_e}^{\rm inj}(t),
\end{equation}
where $n_{p_e}(t) = 4\pi p_e^2 f_{p_e}(t)$ is the number density of electrons per momentum bin, $dp_e / dt = p_e t_{\rm cool}^{-1}$ is the electron cooling rate and $\dot{n}_{p_e}^{\rm inj}(t) \propto p_e^{-s} {\rm exp} \left(-p_e/p_{e, \rm max}\right)$ for $p_e > p_{e, \rm min}$ is the electron injection rate which follows a power-law distribution with exponential cutoff at the maximum momentum $p_{e, \rm max}$. The value of the minimum momentum $p_{e,\rm~min}$ is determined with Eq.~\ref{eq:gamma_min} and the value of $p_{e, \rm max}$ could be determined by $\gamma_{\rm max}$ as mentioned in Eq.~\ref{eq:nu_max}.
The total number of electrons is normalized by matter conservation $N_e = (4\pi / 3) r_s^3 n_{\rm ext} f_e$, where $n_e \approx N_e / (4\pi r_s^2 t_{\rm dyn} \beta_s c)$ is the comoving frame electron number density in the shocked region and $t_{\rm dyn} \approx \Gamma_s t$ is the comoving frame dynamical timescale.
The advantage of using Eq.~\ref{eq:powerlaw} is that the number of electrons still follows powerlaw distribution even in the deep Newtonian phase when $\gamma_m < 2$.
In the deep Newtonian phase, the minimum electron momentum $p_{e, \rm min}$ dominates the total number of electrons, while only electrons with Lorentz factor $\gamma_e \gtrsim 2$ contribute to the synchrotron emission.
In the no escape limit where $t_{\rm esc} = \infty$, the electron distribution can be determined from the following function~\citep[e.g.,][]{dermer_high_2009, murase_implications_2011, zhang_external_2021},
\begin{equation}
   n_{p_e}(t) = \frac{1}{p_e {t^{-1}_{\rm cool}}} \int_{p_e}^\infty dp_e^\prime \dot{n}_{p_e^\prime}^{\rm inj}(t),
\end{equation}
where $t^{-1}_{\rm cool} = t^{-1}_{\rm syn} + t^{-1}_{\rm IC} +t^{-1}_{\rm ad}$ including synchroton cooling, inverse-Compton cooling and adiabatic cooling, respectively.
Once we get the non-thermal electron distribution, we can calculate the synchrotron emission.
The effect of SSA and FF is self-consistently considered in the numerical calculations.

\subsection{Comparison between the analytical and numerical method}

\begin{figure}
	\subfigure[]{
		\includegraphics[scale=0.6]{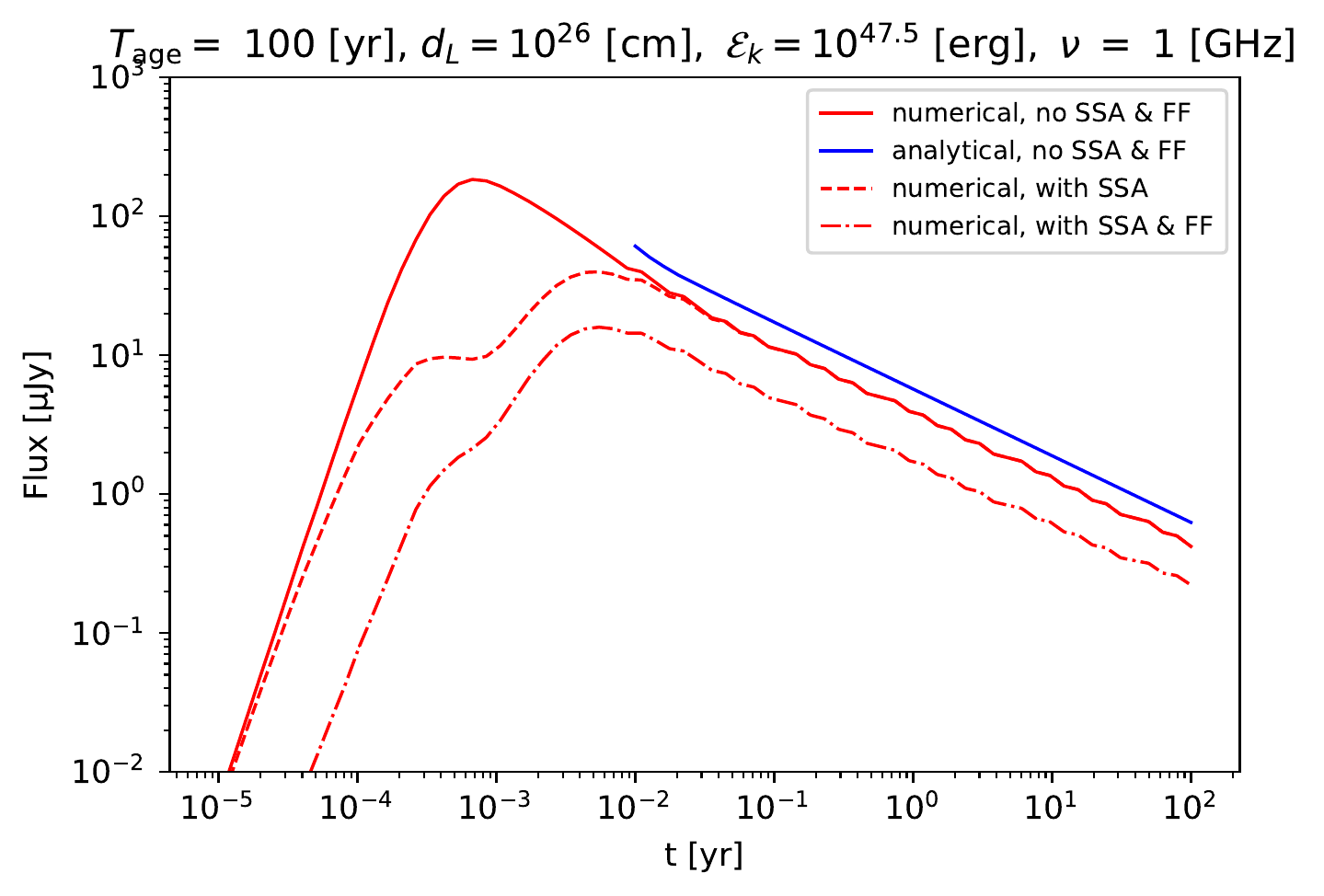}
		\label{fig:flux-low}
	}
    \caption{Light curves at $\nu = 1 {\rm~GHz}$ calculated with both analytical (blue lines) and numerical (red lines) methods for an external baryonic density of $n_{\rm ext} \simeq 2900 \rm~cm^{-3}$. The solid lines represent the results without considering SSA and FF absorption. The dashed line considers the effect of SSA absorption only. In the dash-dotted line, both SSA and FF absorption are taken into account.
    The microphysical parameters used here are $s = 2.2$, $f_e = 0.05$, $\epsilon_e = 0.01$, and $\epsilon_B = 0.001$.
     }
	\label{fig:ananum_flux}
\end{figure}

We compare the resulting synchrotron light curves at $\nu = 1 {\rm~GHz}$ calculated with both analytical and numerical methods for the thin-shell approximation in Fig.~\ref{fig:ananum_flux}.
For demonstrative purposes, we assume the kinetic energy is $\mathcal{E}_k = 10^{47.5}\rm~erg$ with a luminosity distance of $d_L = 10^{26}\rm~cm$ and an initial Lorentz factor of $\Gamma_0 = 2$.

In Fig.~\ref{fig:ananum_flux}, the light curves are calculated with $T_{\rm age} = 100 {\rm~yr}$ where the corresponding external baryonic number density is $n_{\rm ext} \simeq 2900 {\rm~cm^{-3}}$.
We can see the consistency of the flux at deceleration phase between numerical (red-solid line) and analytical (blue-solid line) light curves without absorption. We do not show the analytical light curves at the early times since the calculation of analytical method in early time is not accurate.
Note that when $t \approx 1.7 \times 10^{-2} {\rm~yr}$, the shock velocity $\beta_s$ becomes smaller than $\beta_{\rm DN} \approx 0.18$.
We show the effect of both FF absorption and SSA for the numerical light curve (red dash-dotted line), which is important at the early times.
When the density of the external baryonic matter becomes larger, which usually occurred at smaller $T_{\rm age}$, the FF absorption would be more significant.

\section{Results}
\label{sect:results}
\subsection{Light curves and spectra}
\label{sect:M81_lightcurve}

We consider bursts with outburst energy of $\mathcal{E}_k=10^{42}-10^{47}$~erg and assume that the outflows are largely isotropic. 
For observed GRBs that presumably originate from MGFs, such as GRB 200415A, according to the summary of \cite{Burns_2021ApJ...907L..28B}, the range of their isotropic energy is from a few $10^{44} {\rm~erg}$ to a few $10^{46} {\rm~erg}$. Considering the relationship between kinetic energy and isotropic energy, the energy interval we choose is reasonable. The quasi-isotropic assumption is also reasonable for MGFs although the isotropic-equivalent energy can be larger if beamed. In the following sections, we present our results based on numerical calculations with $s = 2.2$ and $\Gamma_0 = 2$.

\cite{Bhardwaj_2021ApJ...910L..18B} reported a repeating FRB 20200120E and they suggested that M 81, which is a spiral galaxy at $3.63 \pm 0.34 \rm~Mpc$, is the most likely host of FRB 20200120E. We then consider there is a burst that occurred at $d_L = 4 {\rm~Mpc}$ which is similar to the luminosity distance of galaxy M 81.

\subsubsection{Thin shell}
\begin{figure*}
	\centering
	\subfigure[]{
		\includegraphics[scale=0.6]{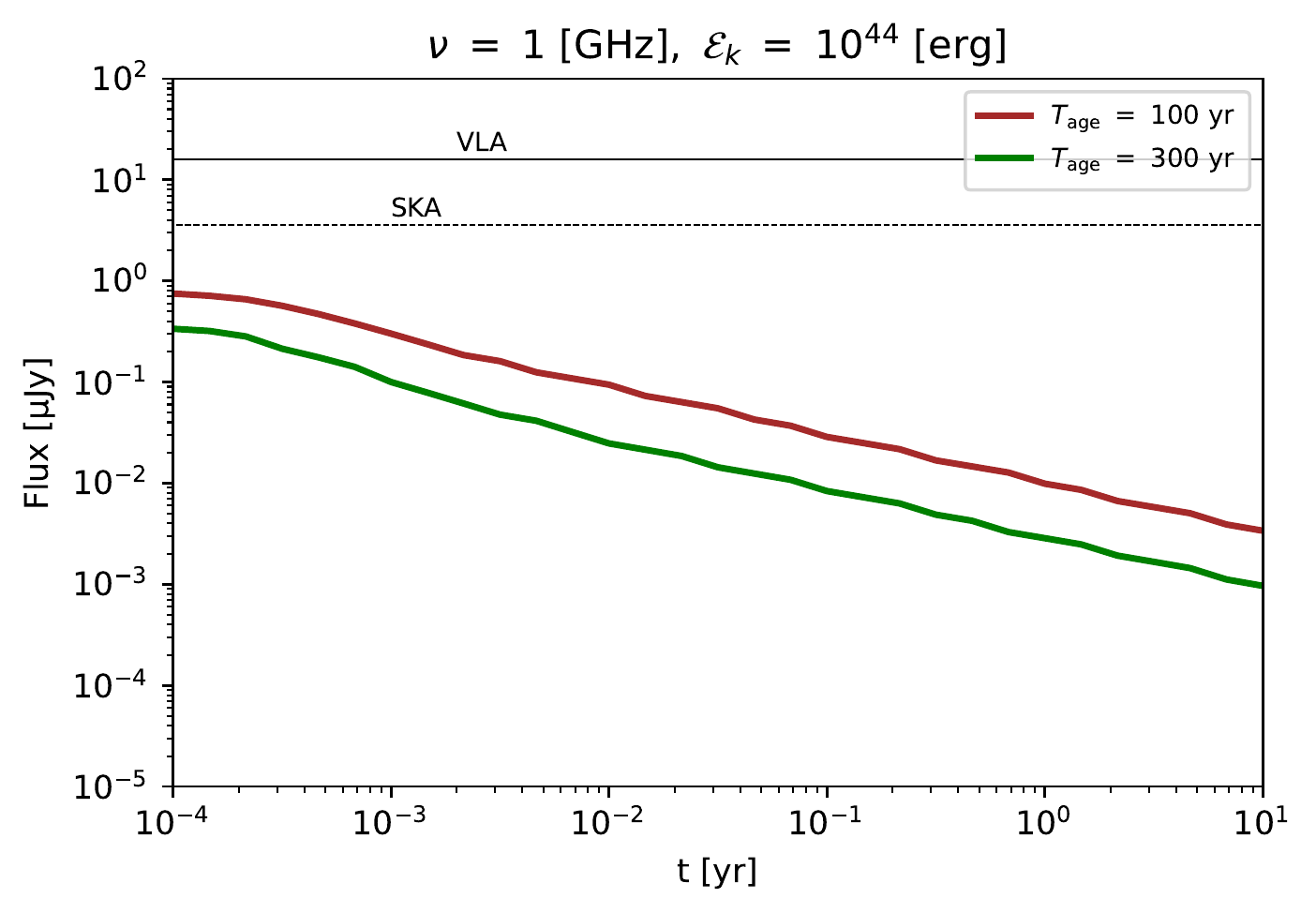}
		\label{fig:light curve_M81_1e44_1GHz}
	}%
	\subfigure[]{
		\includegraphics[scale=0.6]{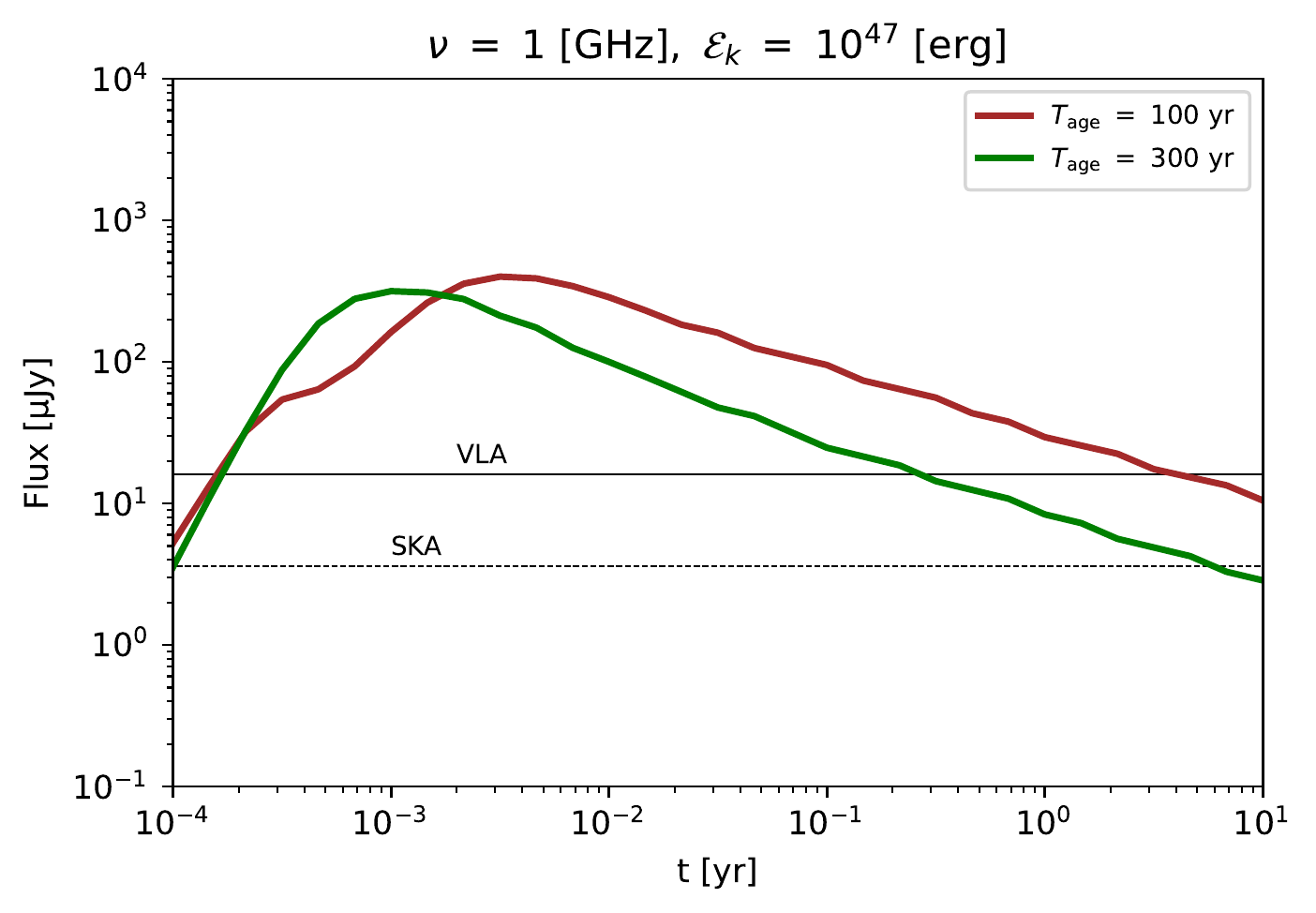}
		\label{fig:light curve_M81_1e47_1GHz}
	}%
	
	\subfigure[]{
		\includegraphics[scale=0.6]{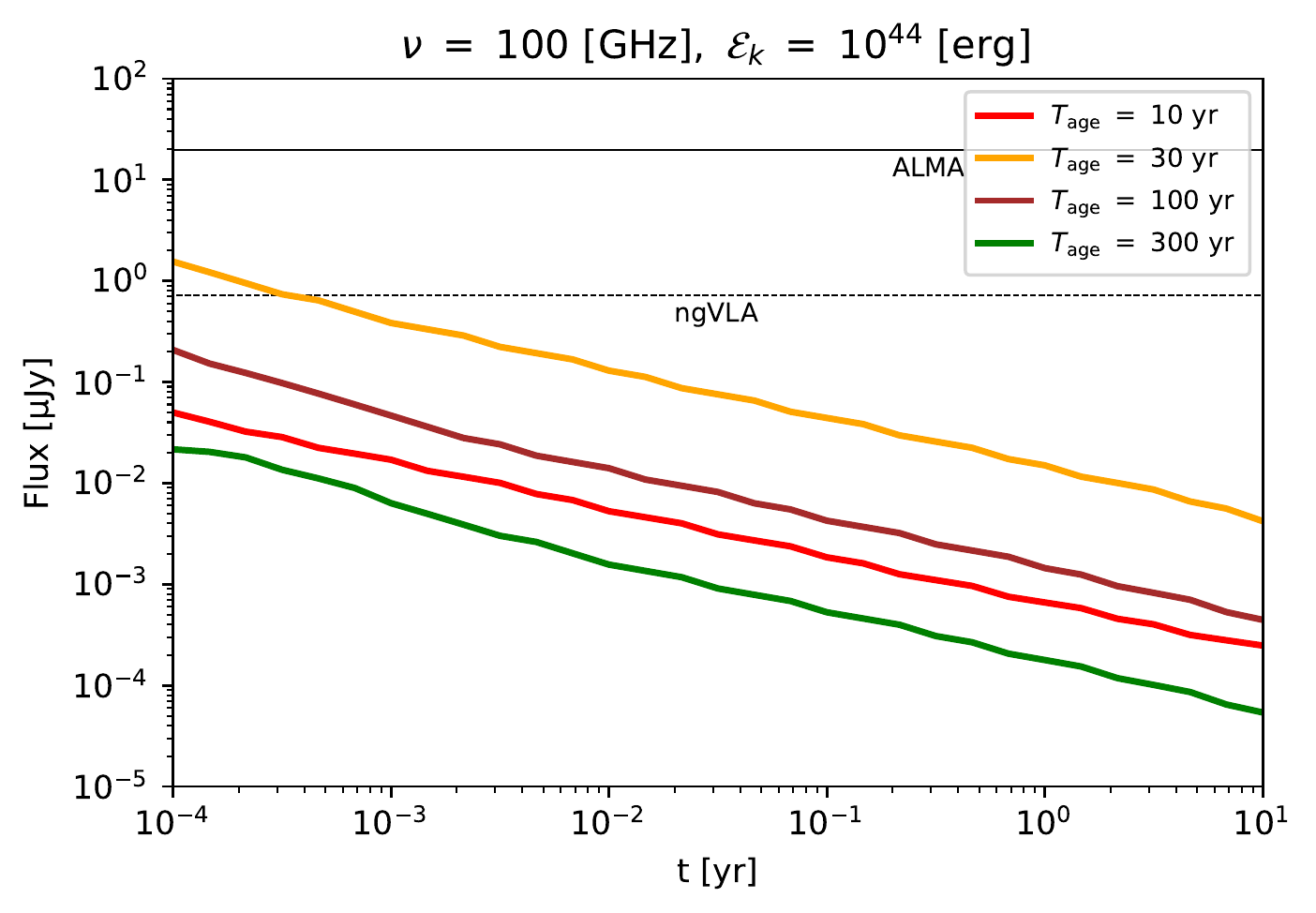}
		\label{fig:light curve_M81_1e44_100GHz}
	}%
	\subfigure[]{
		\includegraphics[scale=0.6]{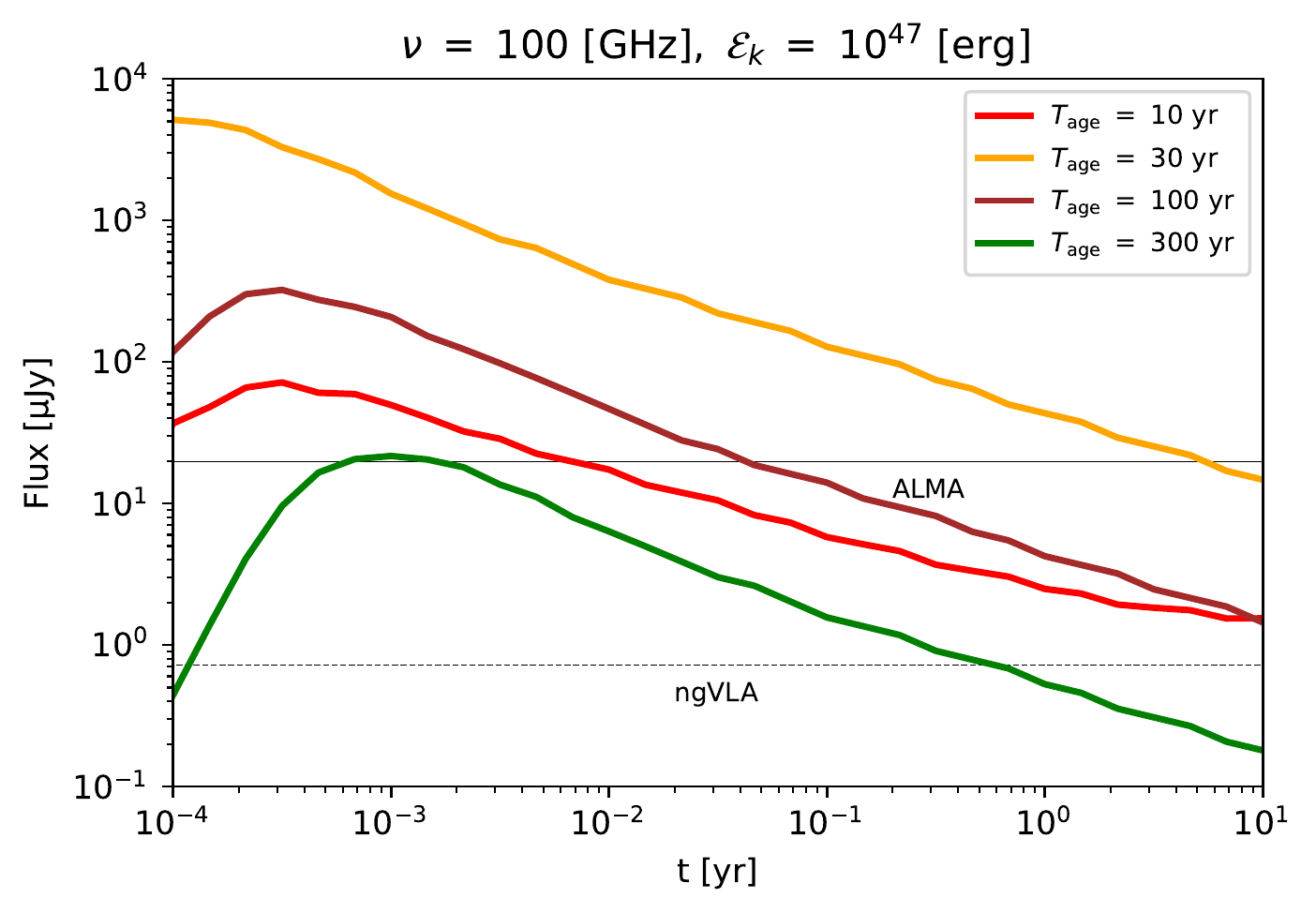}
		\label{fig:light curve_M81_1e47_100GHz}
	}%
		
	\caption{Light curves of afterglow emission at 1 GHz (upper panels) and 100 GHz (lower panels), respectively. 
	The light curves are calculated for different values of $T_{\rm age}$, as indicated in the caption. 
	We show the sensitivity curves of VLA (solid line) and SKA (dashed line) at 1 GHz in the upper panels and those of ALMA (solid line) and ngVLA (dashed line) at 100 GHz in the lower panels.
	We show the light curves with $\mathcal{E}_k = 10^{44} {\rm~erg}$ (left panels) and $\mathcal{E}_k = 10^{47} {\rm~erg}$ (right panels), respectively. Note that the light curves at the 1 GHz band are significantly reduced when $T_{\rm age} \lesssim 30 \rm~yr$ due to strong FF absorption, which are not shown in the upper panels.
	The corresponding physical parameters are $\Gamma_0 = 2$, $d_L = 4 {\rm~Mpc}$, $s = 2.2$, $f_e = 0.05$, $\epsilon_e = 0.01$ and $\epsilon_B = 0.001$.
	}
	\label{fig:light curve_M81}
\end{figure*}

In Fig.~\ref{fig:light curve_M81}, we show the light curves of the observed flux at 1 GHz and 100 GHz, respectively.
The kinetic energy of the burst is assumed to be $\mathcal{E}_k = 10^{44} {\rm~erg}$. 
For comparison, we also consider the case with $\mathcal{E}_k = 10^{47} {\rm~erg}$ even though it is rare for bursts related to MGFs and/or FRBs.
As shown in the previous section, the external baryonic ejecta density has a significant effect on the observed flux due to the SSA and FF effect, and it is determined by the mass of the SN ejecta $M_{\rm ext}$ and the age of the magnetar $T_{\rm age}$ as shown in Eq.~\ref{eq:n_ext}.
In this work, we fix $M_{\rm ext} = 10 M_\odot$, which is reasonable for SN ejecta. Considering the uncertainty of the onset time of the MGFs and/or FRBs since SN explosion, we consider $T_{\rm age} = 10\rm~yr, 30\rm~yr, 100\rm~yr$ and $300\rm~yr$, respectively. Note that the shock will turn into the radiative phase when $t = t_{\rm cool}$~\citep[e.g.,][]{Blondin_1998ApJ...500..342B}, where $t_{\rm cool} \simeq 0.69 k T_{\rm sh}/ (n_{\rm ext} \Lambda_{\rm cool})$ is the cooling time, $\Lambda_{\rm cool} \sim 10^{-16} {~\rm erg ~ cm^3 ~ s^{-1}} \ T_{\rm sh}^{-1}$ is the temperature-dependent volume cooling function, $T_{\rm sh} \simeq 1.9 \times 10^{10} {~\rm K} \ \mathcal{E}_{k, 47}^{2/5} n_{\rm ext, 3}^{-2/5} t_6^{-6/5}$ is the temperature of the shock-heated gas for an adiabatic shock. Noting that the shock crossing time is comparable to the dynamical time $t_{\rm dyn} \approx r_s / v_s$, the transition time from the adiabatic phase to the radiative phase is $t_{\rm tr} \simeq 59 {~\rm yr} \ \mathcal{E}_{k, 47}^{4/17} n_{\rm ext, 3}^{-9/17}$. For $T_{\rm age} = 10 - 100 \rm~yr$, the radiative transition time is $\sim 1 - 10 \rm~yr$. Therefore, we just show the light curves at $t \leq 10 ~\rm yr$.

The sensitivity of current and next-generation radio telescopes are considered\footnote{The official website of these radio telescopes: VLA (http://www.vla.nrao.edu), SKA (https://www.skatelescope.org), ALMA (https://public.nrao.edu/telescopes/alma/) and ngVLA (https://ngvla.nrao.edu/).}. 
For VLA and ALMA, the root-mean-square (RMS) sensitivities could be obtained using the calculators~\footnote{The calculator website of VLA: https://obs.vla.nrao.edu/ect, and that of ALMA: https://almascience.nrao.edu/proposing/sensitivity-calculator.} with exposure time $t_{\rm exp} = 10^4 {~\rm s}$. 
For SKA and ngVLA, we found the RMS sensitivities with $t_{\rm exp} = 3600 {~\rm s}$ from their official websites, and the corresponding RMS with $t_{\rm exp} = 10^4 {~\rm s}$ could be derived using the relation ${\rm RMS} \propto t_{\rm exp}^{-1/2}$.
The $3 \sigma$ upper limits $S = 3~{\rm RMS}$ are indicated as black lines in Fig.~\ref{fig:light curve_M81}, which are $16.0 \rm{~\mu Jy}$ (for VLA at 1~GHz), $3.6 \rm{~\mu Jy}$ (for SKA at 1~GHz), $19.7 ~\rm{\mu Jy}$ (for ALMA at 100~GHz) and $0.72 \rm{~\mu Jy}$ (for ngVLA at 100~GHz), respectively.

In Fig.~\ref{fig:light curve_M81_1e44_1GHz} and Fig.~\ref{fig:light curve_M81_1e44_100GHz}, we show the light curve with $\mathcal{E}_k = 10^{44} {\rm~erg}$.
We can see that the afterglow emission with $T_{\rm age} = 10 {\rm~yr}$ and $30 {\rm~yr}$ at $\nu = 1 {\rm~GHz}$ is almost reduced to zero since the FF absorption is strong when $T_{\rm age}$ becomes smaller. 
Due to the smaller value of the burst kinetic energy, the detection is still difficult even with larger $T_{\rm age}$ where the FF absorption could be neglected.
However, the detection of the afterglow emission for ngVLA at $\nu = 100 {\rm~GHz}$ is possible at earlier time $t \lesssim 1.0 \times 10^{4} {\rm~s}$ if $T_{\rm age} = 30 {~\rm yr}$ where the effect of FF absorption is not important at higher frequency band. 
In Fig.~\ref{fig:light curve_M81_1e47_1GHz} and Fig.~\ref{fig:light curve_M81_1e47_100GHz}, we show the light curve with $\mathcal{E}_k = 10^{47} {\rm~erg}$. 
Different from the previous case, both VLA and SKA could detect the afterglow emission with $T_{\rm age} = 100 {\rm~yr}$ and $300 {\rm~yr}$ for $\nu = 1 {\rm~GHz}$. 
For $\nu = 100 {\rm~GHz}$, the afterglow with $T_{\rm age} = 10 - 300 {\rm~yr}$ could be detected by ngVLA, but for ALMA detecting the afterglow with $T_{\rm age} = 300 {\rm~yr}$ is difficult.
Note that the increase of the light curve in Fig.~\ref{fig:light curve_M81_1e47_100GHz} with $T_{\rm age} = 10 {\rm~yr}$ (red solid line) in late time is due to the gradual weakening of FF absorption when the forward shock propagates near the edge of the external baryonic ejecta (that is $r_s / R_{\rm ext} \sim 30 \% $ ).

\begin{figure*}
	\subfigure[]{
		\includegraphics[scale=0.6]{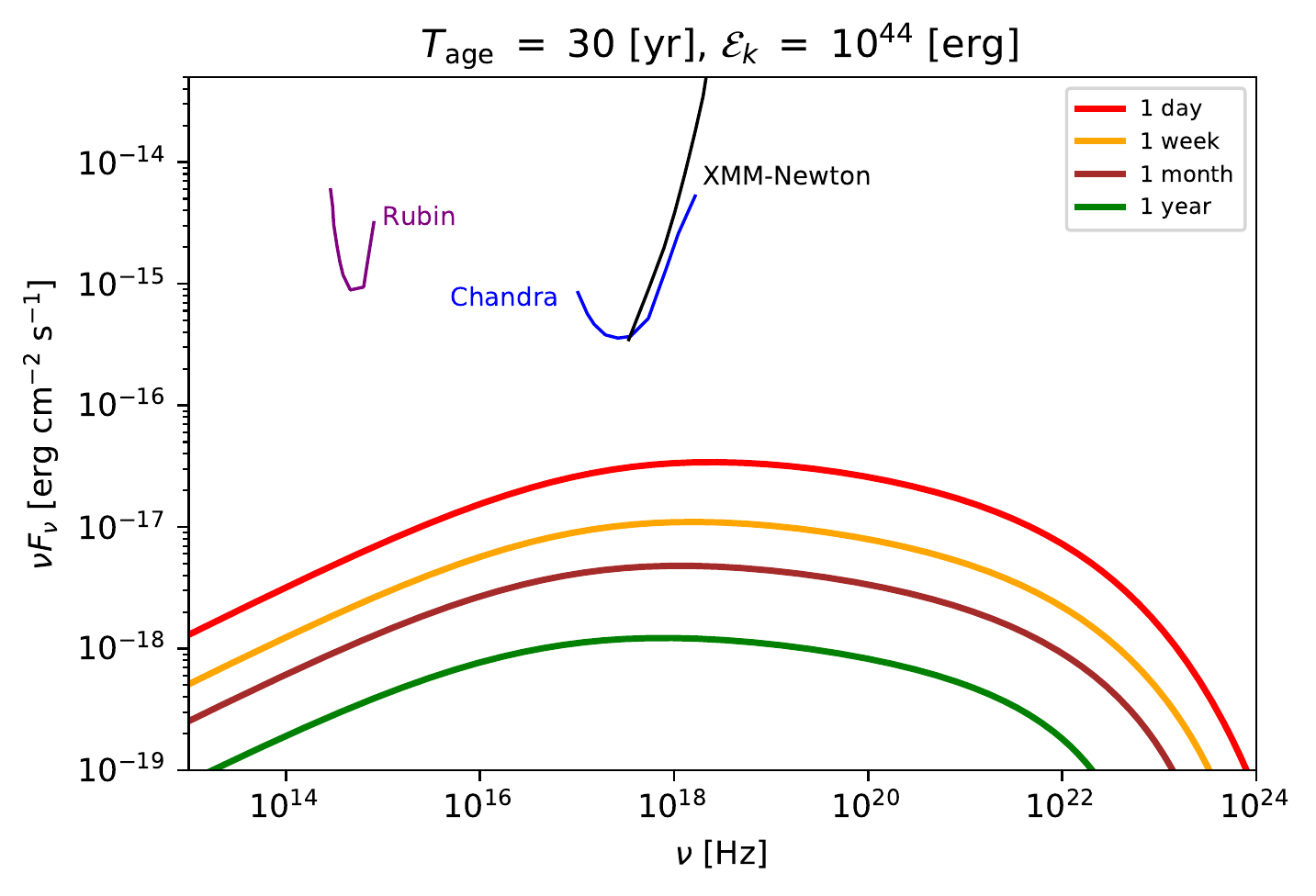}
	}%
	\subfigure[]{
		\includegraphics[scale=0.6]{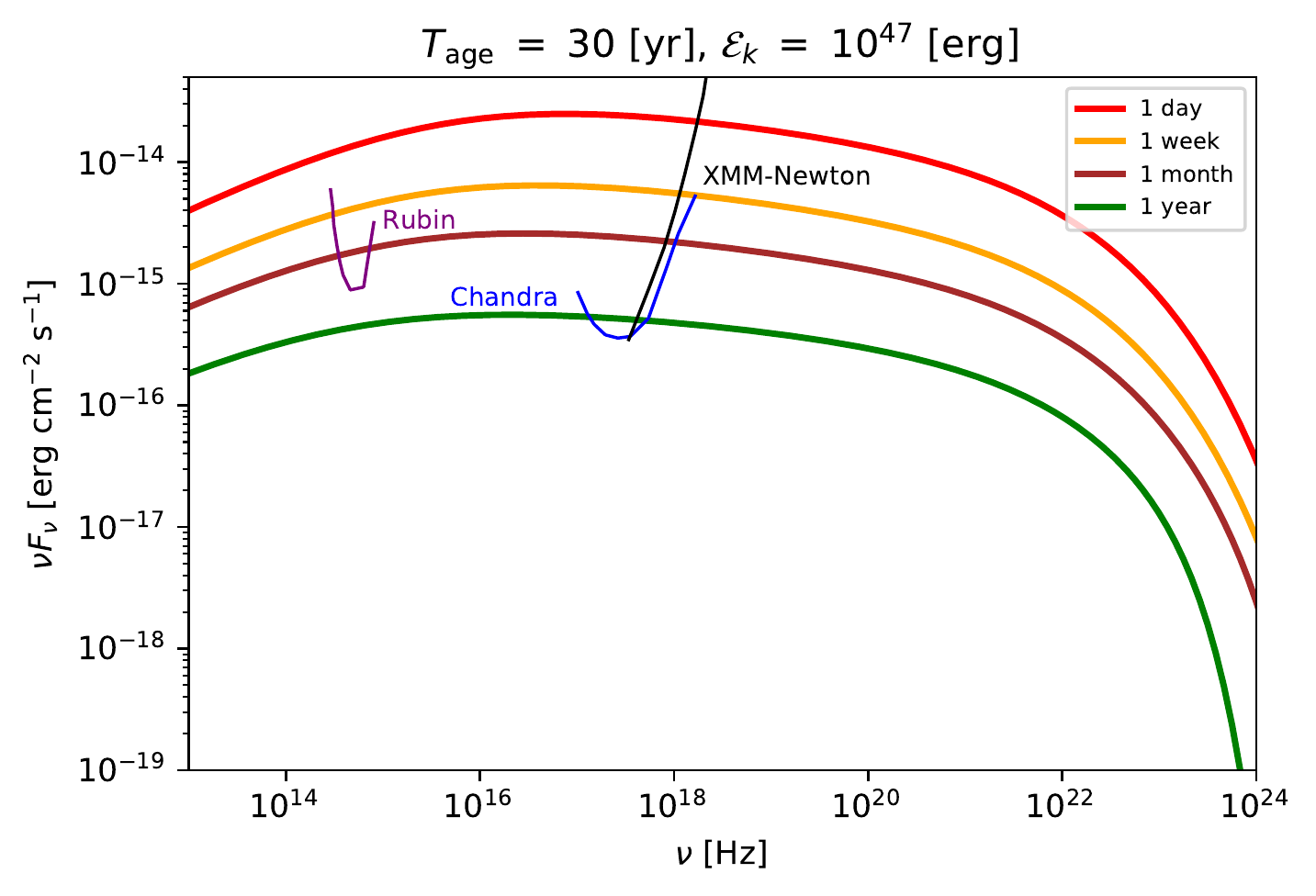}
	}%

	\subfigure[]{
		\includegraphics[scale=0.6]{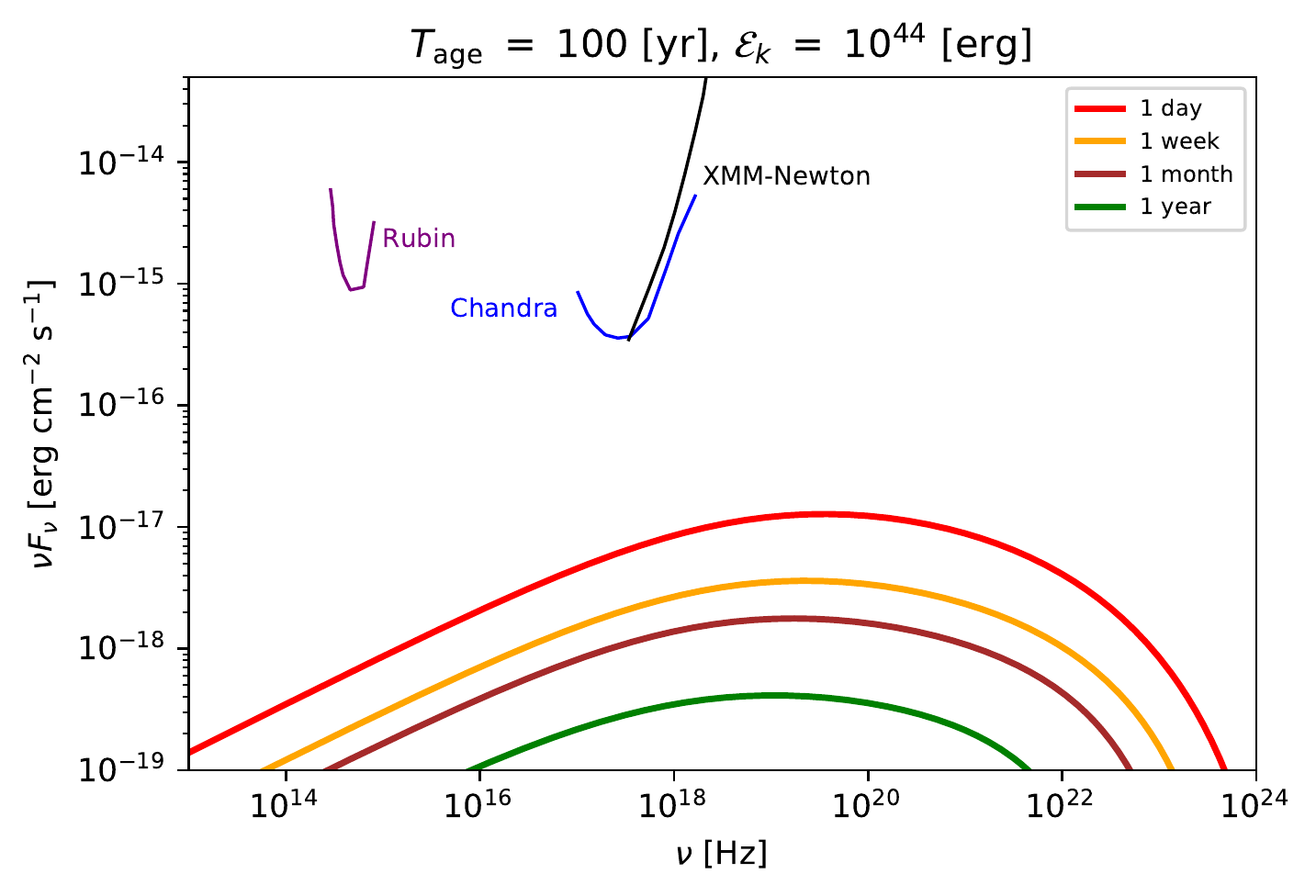}
	}%
	\subfigure[]{
		\includegraphics[scale=0.6]{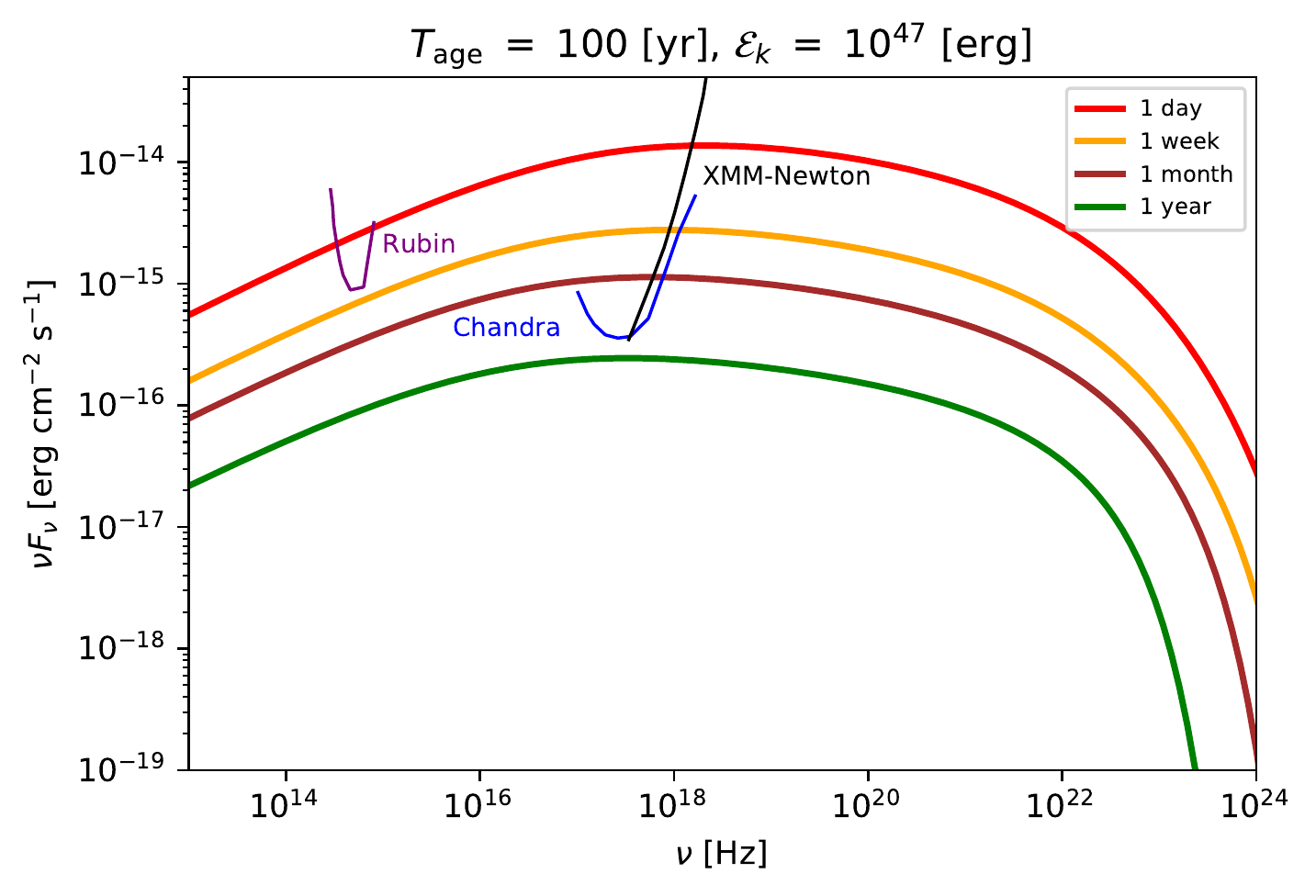}
	}%
	\caption{Energy spectra of the afterglow emission predicted in the burst-in-bubble model for $\mathcal{E}_k = 10^{44} \rm~erg$ (left panel) and $\mathcal{E}_k = 10^{47} \rm~erg$ (right panel), respectively. The sensitivity curves of \textit{Chandra} (blue line) and \textit{XMM-Newton} (black line) are calculated with an exposure time of $10^5~{\rm~s}$ (\citealt{Lucchetta_2022JCAP}). The purple lines show the r-band sensitivity of the Vera C. Rubin Observatory with a point source exposure time of $30 {\rm~s}$ in the 3-day revisit time (\citealt{Yuan_2021ApJ...911L..15Y}). In the upper panels and lower panels, we set $T_{\rm age}$ to be 30~yr and 100~yr, respectively. The corresponding physical parameters are $\Gamma_0 = 2$, $d_L = 4 {\rm~Mpc}$, $s = 2.2$, $f_e = 0.05$, $\epsilon_e = 0.01$ and $\epsilon_B = 0.001$.}
	\label{fig:spectrum_M81_cgs}
\end{figure*}

In Fig.~\ref{fig:spectrum_M81_cgs}, we show the energy spectra of the afterglow emission for $\mathcal{E}_k = 10^{44}\rm~erg$ (left panel) and $\mathcal{E}_k = 10^{47}\rm~erg$ (right panel) at different observation times from one day to one year. 
In the upper panels and lower panels, we show the spectra with $T_{\rm age} = 30 {\rm~yr}$ and $T_{\rm age} = 100 {\rm~yr}$, respectively.
At the X-ray band, we added the sensitivity curves of \textit{Chandra} (blue curve) and \textit{XMM-Newton} (black curve) which are calculated with an exposure time of $10^5~{\rm~s}$.
We can see that the detection of the afterglow emission with $T_{\rm age} = 30 {\rm~yr}$ ($100 {\rm~yr}$) at the X-ray band is promising for $\mathcal{E}_k = 10^{47}\rm~erg$ when the observation time ranges from 1 day to 1 year (1 month), but the X-ray flux is too low to be detected for $\mathcal{E}_k = 10^{44}\rm~erg$. 
However, the X-ray emission may not be detected since the neutral SN ejecta could absorb the X-ray emission even for $\mathcal{E}_k = 10^{47}\rm~erg$. 
In order to judge the significance of the X-ray absorption, we calculate the column density $N_{\rm ext} \approx (R_{\rm ext} - r_s) n_{\rm ext}$ assuming the SN ejecta are uniformly distributed. 
For $T_{\rm age} = 30 {\rm~yr}$, we can get $N_{\rm ext} \simeq 3.2 \times 10^{22} \rm~cm^{-2}$ ($3.0 \times 10^{22} \rm~cm^{-2}$) for $t = 1 \rm~d$ ($1 \rm~yr$) where the corresponding external baryonic number density is $n_{\rm ext} \simeq 1.1 \times 10^5 {\rm~cm^{-3}}$. 
For hydrogen gas, the effect of X-ray absorption can be neglected if the column density is smaller than $K_X/m_H \sim 6.7 \times 10^{21}~{\rm~cm^{-2}}~{(h\nu/10~\rm keV)}^{-3}$ \citep[e.g.,][]{2015ApJ...805...82M}.
Our results indicate that it is possible for \textit{XMM-Newton} and \textit{Chandra} to detect the afterglow emission in the hard X-ray band.

We also show the sensitivity curve of the Vera C. Rubin Observatory in Fig.~\ref{fig:spectrum_M81_cgs}. 
The detection with the Vera C. Rubin Observatory at the optical band is promising for $\mathcal{E}_k = 10^{47}\rm~erg$ at $t \lesssim 1 {\rm~month}$ ($1 \rm~d$) for $T_{\rm age} = 30 {\rm~yr}$ ($100 {\rm~yr}$). For the fluxes with $\mathcal{E}_k = 10^{44}\rm~erg$, they would be too faint to be detected with the Vera C. Rubin Observatory.

\subsubsection{Thick shell}

The typical value of the nebula width $\Delta R_{\rm ej}$ could be estimated as
\begin{equation}
    \Delta R_{\rm ej} = \frac{R_{\rm nb}}{12 \Gamma_0^2} \simeq 2.1 \times 10^{14} {\rm~cm} \ R_{\rm nb, 16},
\end{equation}
where $R_{\rm nb} \simeq 1.6 \times 10^{16} {~\rm cm} \ P_{i, 0}^{-2/5} M_{\rm ext, 1 M_\odot}^{-1/5} V_{\rm ext, 8.5}^{3/5} T_{\rm age, 10 yr}$ is the radius of nebula \citep[e.g.,][]{Murase:2016sqo}, and $P_i$ is the initial rotation period of magnetar. When the inner ejecta propagates into the SN baryonic ejecta and spreading is negligible, the reverse shock finishes crossing the inner ejecta at the shock crossing time 
\begin{equation}
t_{\rm cross} = \frac{\Delta R_{\rm ej}}{v_0} \simeq 1.3 \times 10^4 {\rm~s} \ R_{\rm nb, 16}.
\end{equation}
If $t_{\rm cross} > t_{\rm dec}$, the thin-shell approximation is not valid, where the observed flux peaks at $t_{\times}\approx t_{\rm cross}$. Here we consider the thick-shell approximation.

\begin{figure*}
	\centering
	\subfigure[]{
		\includegraphics[scale=0.6]{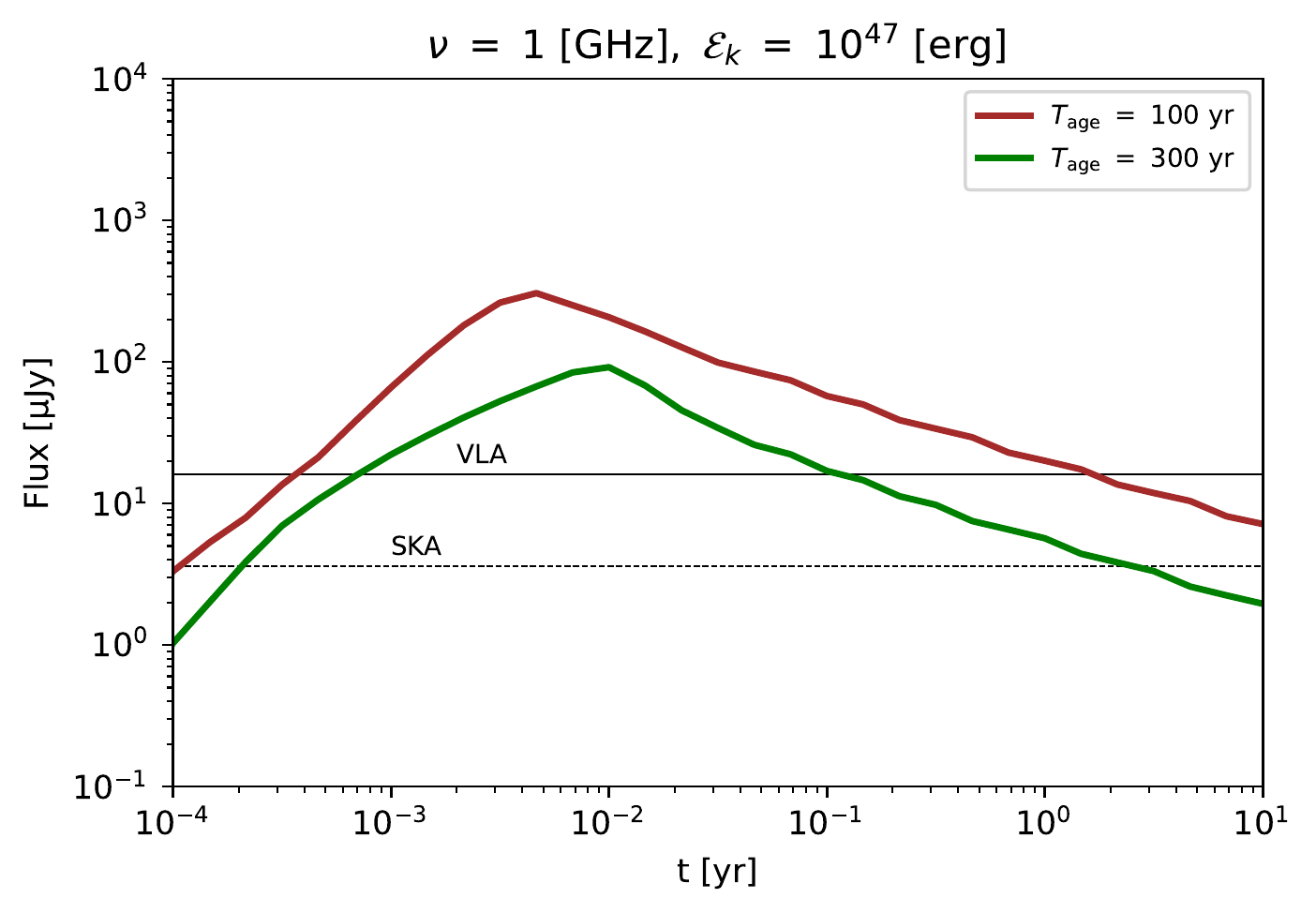}
		\label{fig:light curve_1GHz}
	}%
	\subfigure[]{
	    \includegraphics[scale=0.6]{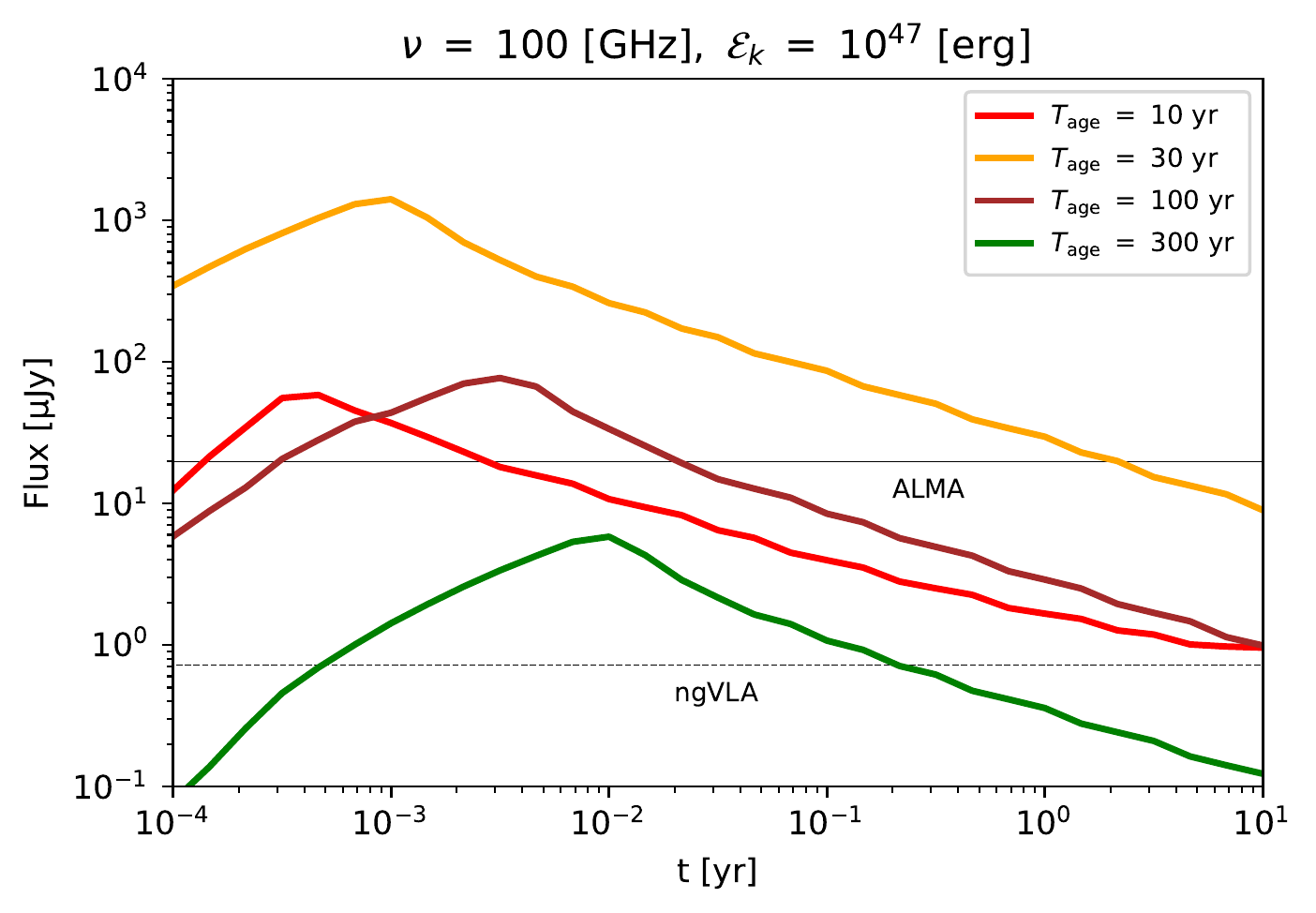}
	    \label{fig:light curve_100GHz}
	}%
	
	\subfigure[]{
		\includegraphics[scale=0.6]{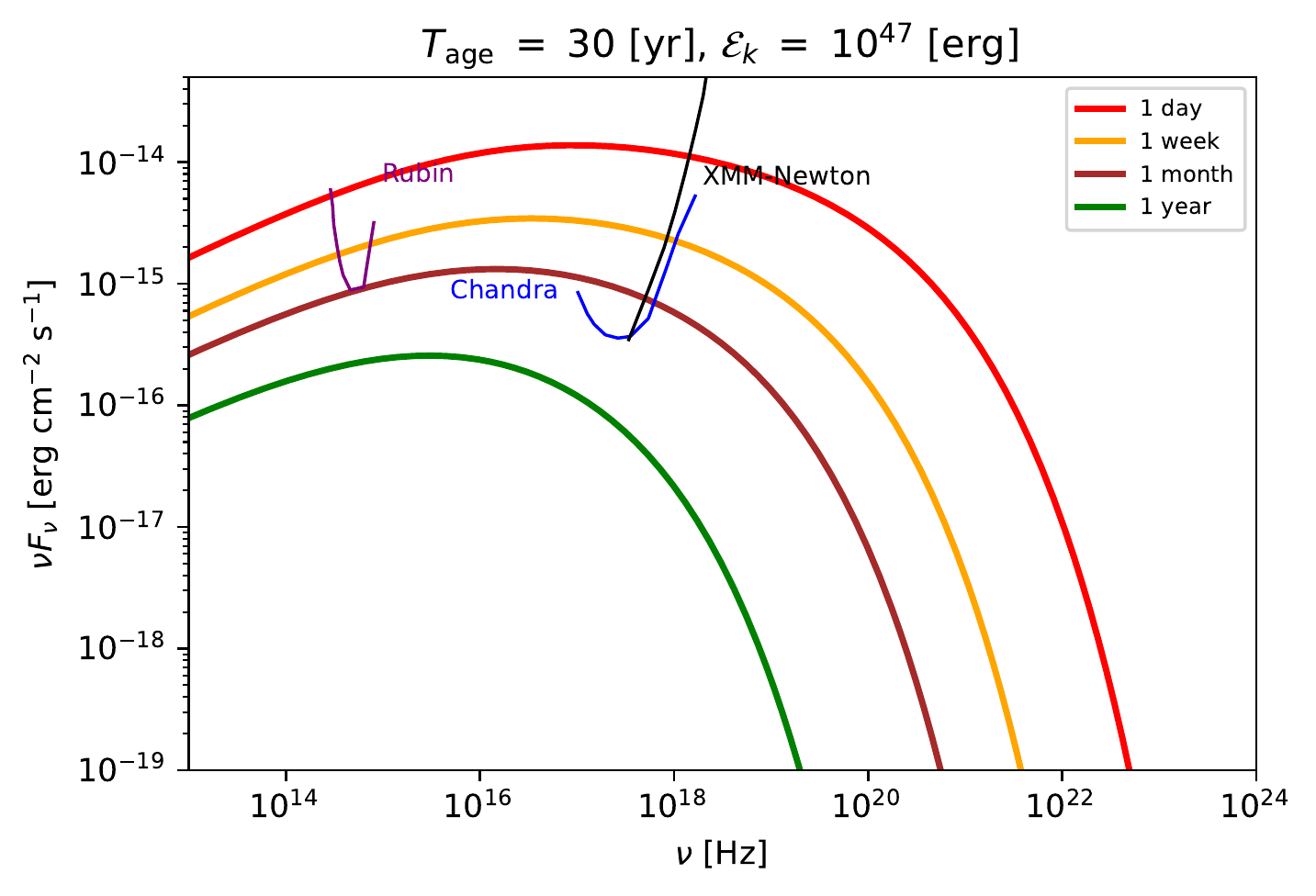}
		\label{fig:spectra_30yr}
	}%
	\subfigure[]{
		\includegraphics[scale=0.6]{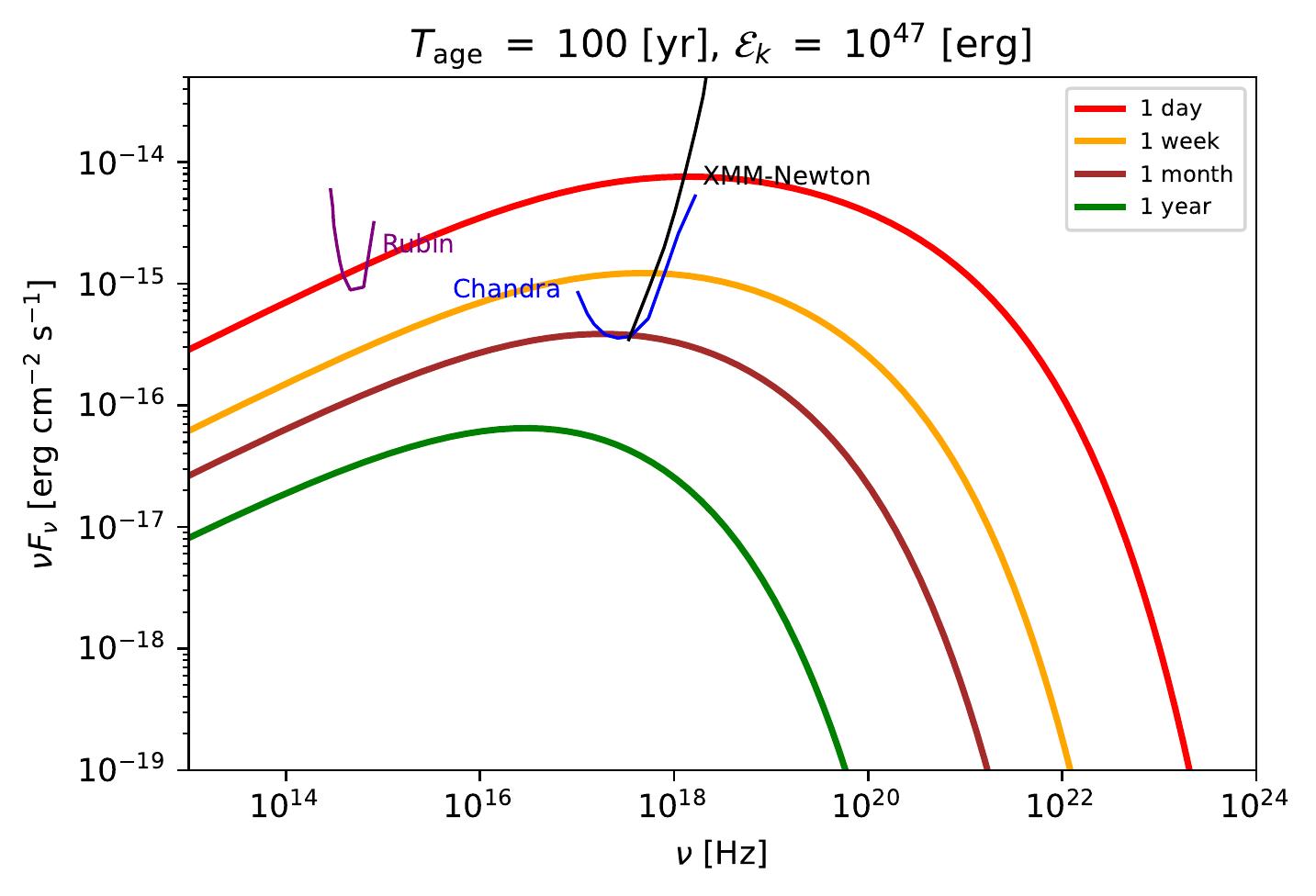}
		\label{fig:spectra_100yr}
	}%
	\caption{Light curves (upper panels) and energy spectra (lower panels) of FRB afterglow emission with $\mathcal{E}_k = 10^{47} ~\rm erg$, respectively. 
	The corresponding physical parameters are $\Gamma_0 = 2$, $d_L = 4 {\rm~Mpc}$, $s = 2.2$, $f_e = 0.05$, $\epsilon_e = 0.01$ and $\epsilon_B = 0.001$.
	}
	\label{fig:thick_splc}
\end{figure*}

In the upper panels of Fig.~\ref{fig:thick_splc}, we show the light curves with $\mathcal{E}_k = 10^{47} ~\rm erg$ at 1 GHz (left panel) and 100 GHz (right panel) for the thick-shell approximation. 
Compared with Figs.~\ref{fig:light curve_M81_1e47_1GHz} and \ref{fig:light curve_M81_1e47_100GHz}, we see that the late-time fluxes are similar. However, the fluxes peak at later times around $t_{\rm cross}$, and the relative fluxes around the peak are lower than those in the thin-shell approximation by a factor of $\sim 1-5$.
For 1 GHz, VLA and SKA can detect the afterglow with $T_{\rm age} = 100-300 ~\rm yr$. For 100 GHz, ALMA (ngVLA) can detect the afterglow with $T_{\rm age} = 10-100 ~\rm yr$ ($T_{\rm age} = 10 - 300 ~\rm yr$).

In the lower panels of Fig.~\ref{fig:thick_splc}, we show the spectra with $\mathcal{E}_k = 10^{47} ~\rm erg$ for $T_{\rm age} = 30 ~\rm yr$ (left panel) and $T_{\rm age} = 100 ~\rm yr$ (right panel). At the X-ray band, \textit{XMM-Newton} and \textit{Chandra} may detect afterglow emission with $t \lesssim 1 ~\rm month$ for $T_{\rm age} = 30 ~\rm yr$ and $T_{\rm age} = 100 ~\rm yr$. At the optical band, it is possible for the Vera C. Rubin Observatory to detect afterglow with $t \lesssim 1 ~\rm month$ ($t \lesssim 1 ~\rm d$) for $T_{\rm age} = 30 ~\rm yr$ ($T_{\rm age} = 100 ~\rm yr$).

\subsection{Detection horizon for radio afterglows}
\label{subsect:2Dplot}
\begin{figure*}
	\centering	 
	\subfigure[]{
		\includegraphics[scale=0.65]{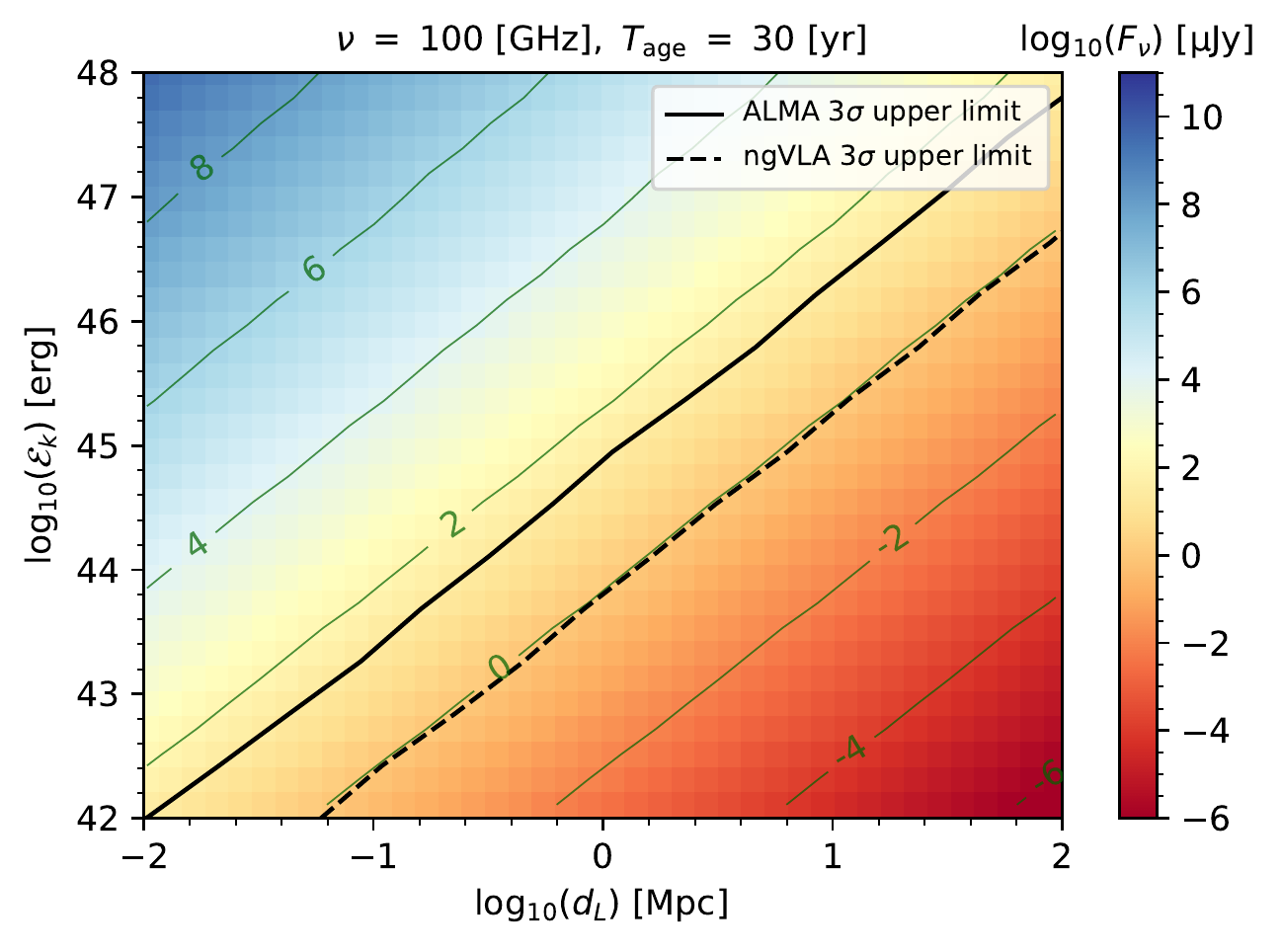}
		\label{fig:2Dplot_100GHz_30yr_0.05}
	}%
	\subfigure[]{
		\includegraphics[scale=0.65]{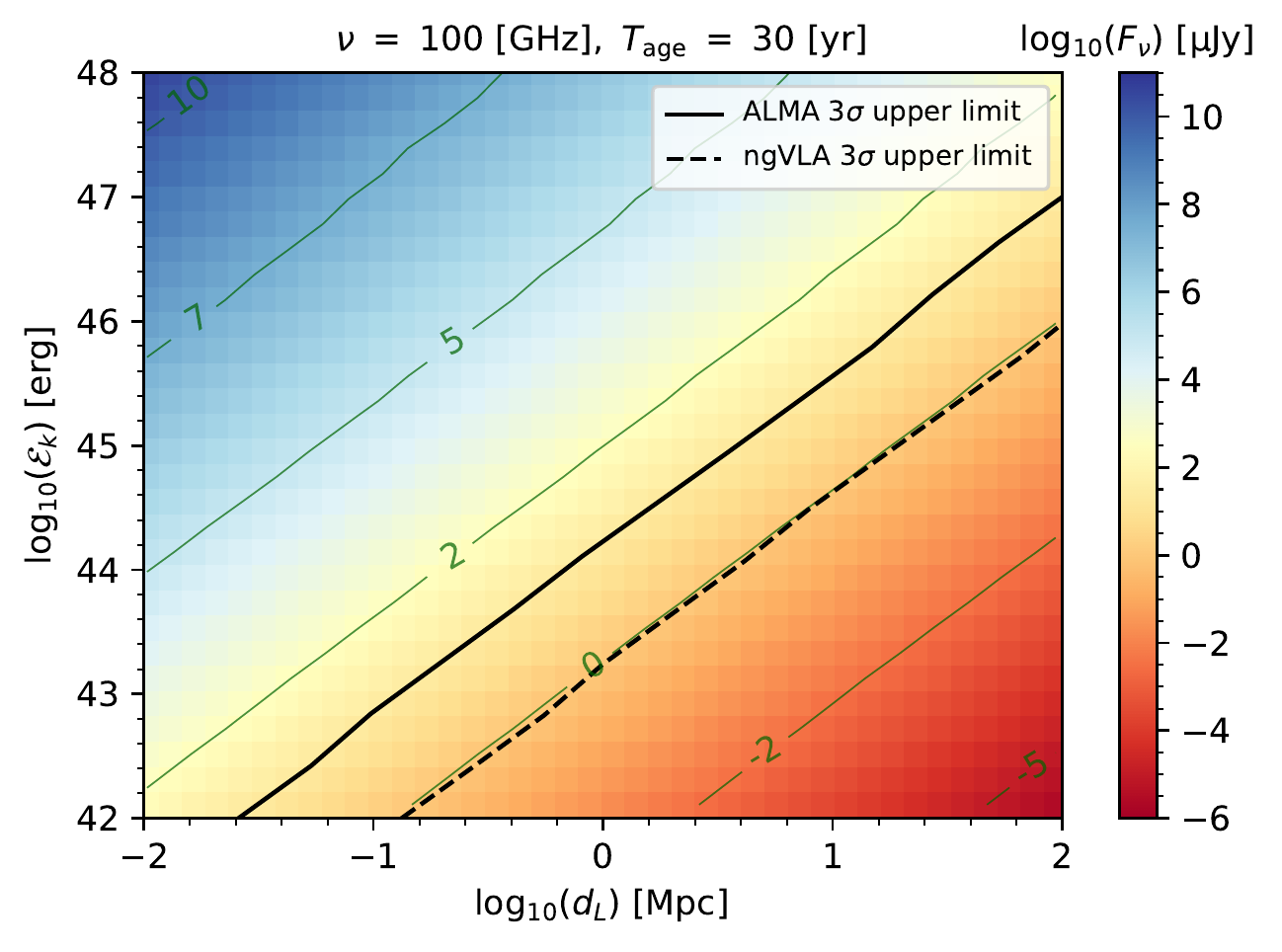}
		\label{fig:2Dplot_100GHz_30yr_0.1}
	}%

	\subfigure[]{
		\includegraphics[scale=0.65]{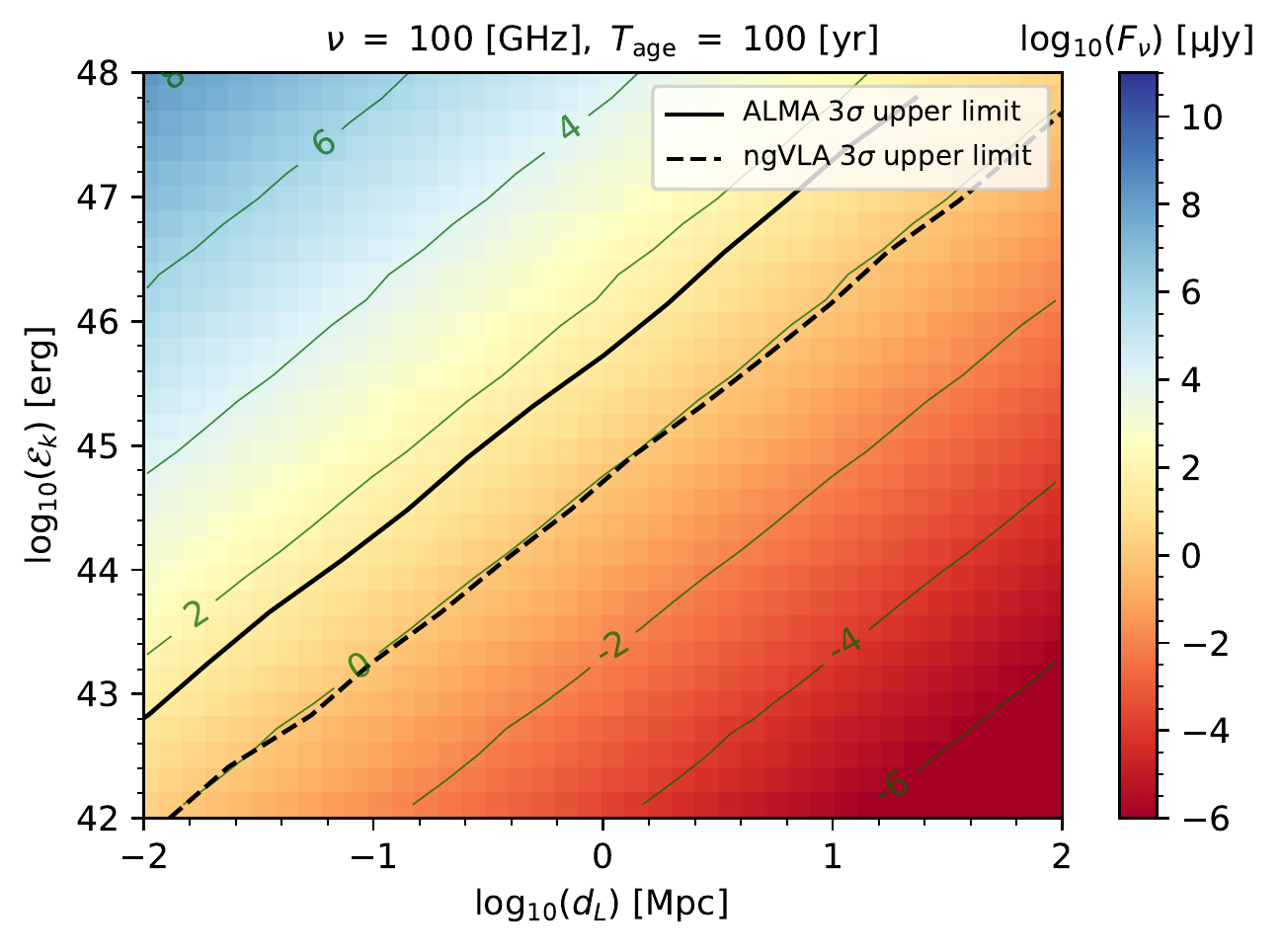}
	}%
	\subfigure[]{
		\includegraphics[scale=0.65]{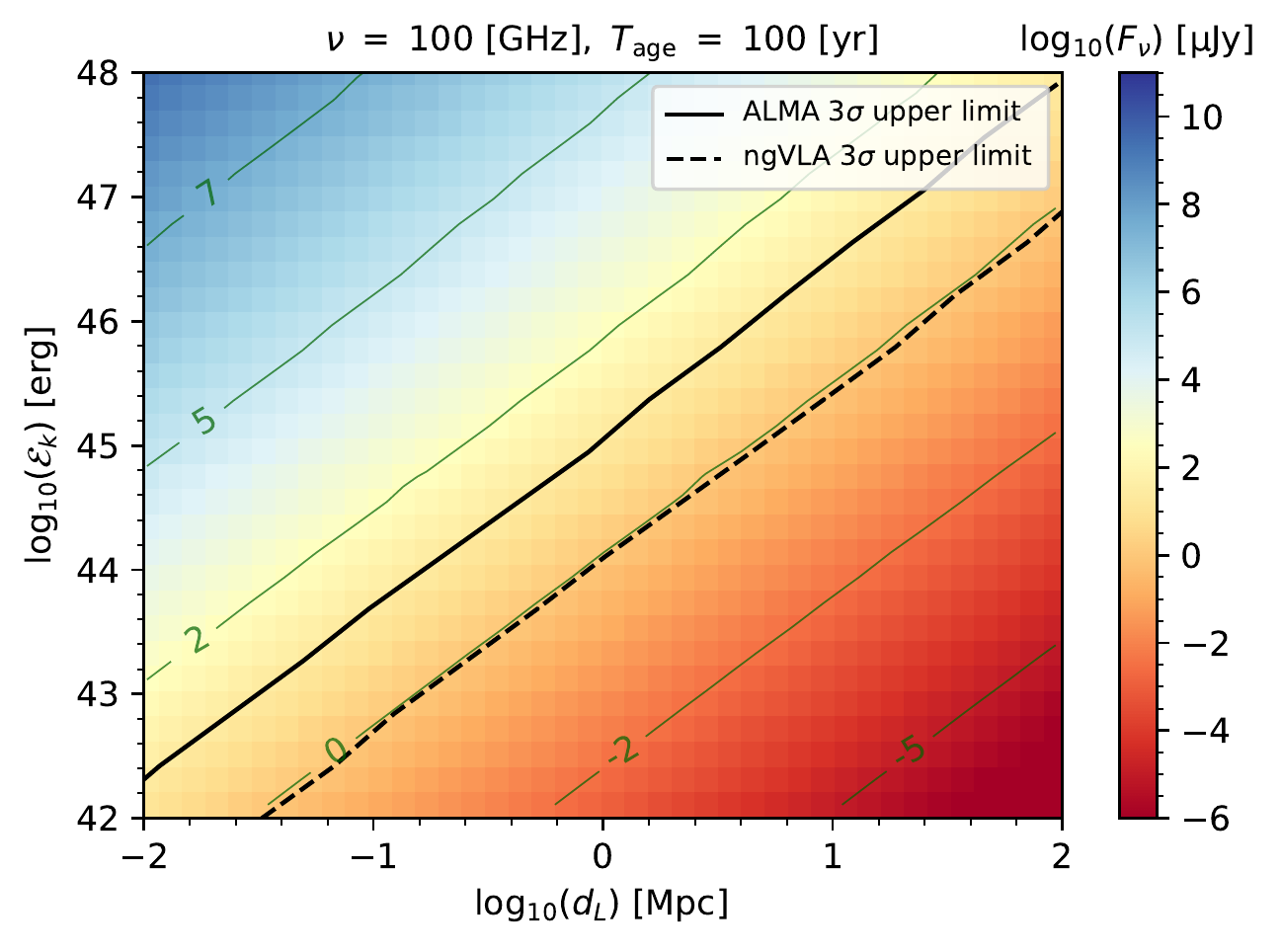}
	}%
	\caption{Parameter space in the $d_{L}-\mathcal{E}_{k}$ plane with the color indicating the observed flux at 100 GHz with $s = 2.2$. The corresponding microphysical parameters are $f_e = 0.05, \ \epsilon_e = 0.01$ and $\epsilon_B = 0.001$ (left panels), and $f_e = 0.1, \ \epsilon_e = 0.1$ and $\epsilon_B = 0.1$ (right panels). The kinetic energy has range $\mathcal{E}_{k} = 10^{42} - 10^{48}~{\rm~erg}$, while the distance ranging from $d_{L} = 10^{-2} - 10^2\rm~Mpc$. In each plot, the green lines are the contour lines in which we write the corresponding value of flux, and the solid and dashed black lines mean the sensitivity of the radio telescopes. In the upper panels and lower panels, we set $T_{\rm age}$ to be $30 {\rm~yr}$ and $100{\rm~yr}$, respectively. }
	\label{fig:2Dplot_100GHz} 
\end{figure*}

\begin{figure*}
	\centering
	\subfigure[]{
		\includegraphics[scale=0.65]{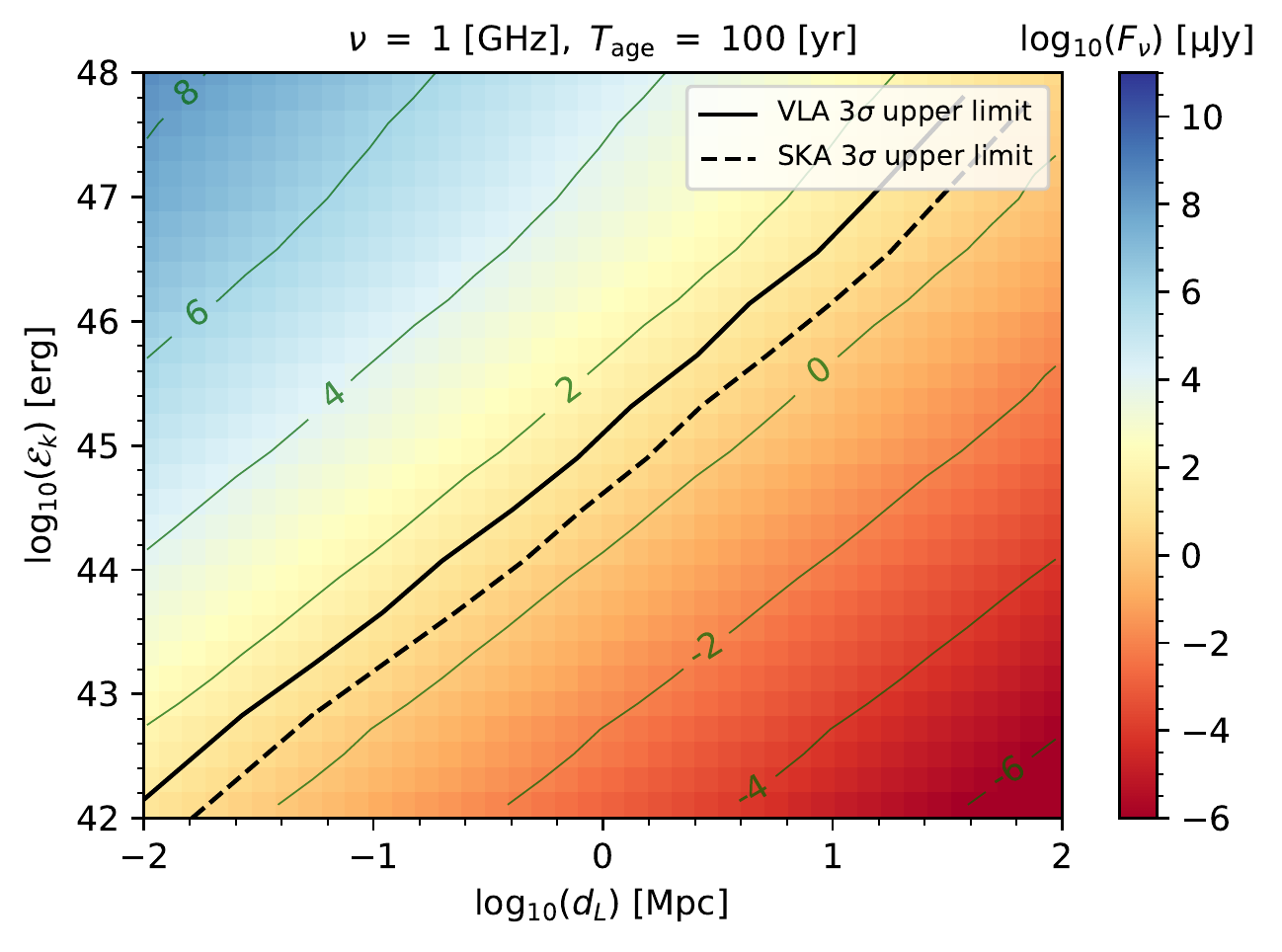}
	}%
	\subfigure[]{
		\includegraphics[scale=0.65]{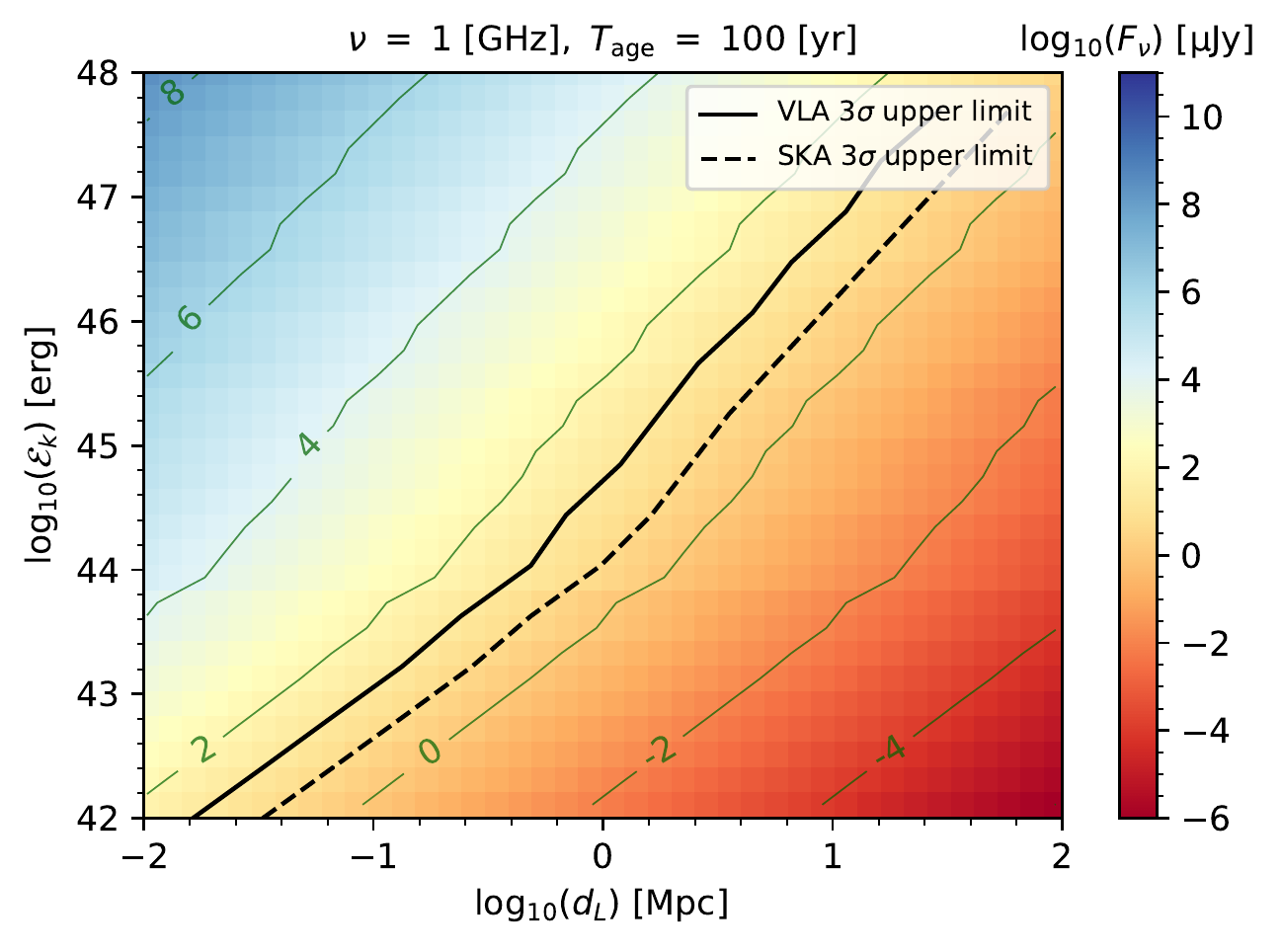}
	}%
		
	\caption{Similar to Fig.~\ref{fig:2Dplot_100GHz}, the color indicates the observed flux at 1 GHz with $s = 2.2$. The corresponding microphysical parameters are $f_e = 0.05, \ \epsilon_e = 0.01$ and $\epsilon_B = 0.001$ (left panel), and $f_e = 0.1, \ \epsilon_e = 0.1$ and $\epsilon_B = 0.1$ (right panel).} 
    \label{fig:2Dplot_1GHz} 
\end{figure*}

In the following sections, we adopt the thick-shell approximation to perform the calculations.

The detectability of afterglow emission is affected by $\mathcal{E}_k$. For a given burst kinetic energy, the detection horizon $d_{L,\rm lim}$ considering the sensitivity curves of radio telescopes is given by
\begin{equation}\label{eq:fittedLINE}
d_{L,\rm lim}={\left(\frac{\nu L_{\nu} (t_{\rm peak})}{4\pi F_{\rm lim}}\right)}^{1/2},
\end{equation}
where $\nu L_{\nu} (t_{\rm peak})$ is the observed luminosity at $t = t_{\rm peak}$, $t_{\rm peak}$ is the time when the luminosity reaches maximum value during the time interval from $10^{-3.5}{\rm~yr}$ to $10^2 {\rm~yr}$, and $F_{\rm lim}$ is the detector sensitivity at a given frequency.  
In Fig.~\ref{fig:2Dplot_100GHz}, the detection horizon at 100 GHz for ALMA (solid curve) and ngVLA (dashed curve) are indicated as black curves in the $d_{L}-\mathcal{E}_{k}$ plane where the color indicates the flux observed at $t_{\rm peak}$, respectively. The green contours are the corresponding values of the logarithm of the observed fluxes.

In the left panels, we show the results for $f_e = 0.05$, $\epsilon_e = 0.01$, and $\epsilon_B = 0.001$, which can be regarded as the conservative case. 
We see that the corresponding detection horizon is more extensive for $T_{\rm age} = 30 {\rm~yr}$ than $T_{\rm age} = 100 {\rm~yr}$ for both ALMA and ngVLA. The detection horizon is larger for ngVLA compared to ALMA. 
For example, in Fig.~\ref{fig:2Dplot_100GHz_30yr_0.05}, the values of $d_{L, \rm lim}$ for $T_{\rm age} = 30 {\rm~yr}$ are $0.01 {\rm~Mpc}$ ($0.06 {\rm~Mpc}$) for $\mathcal{E}_k = 10^{42} {\rm~erg}$, $1.2 {\rm~Mpc}$ ($6.7 {\rm~Mpc}$) for $\mathcal{E}_k = 10^{45} {\rm~erg}$ and $59 {\rm~Mpc}$ ($310 {\rm~Mpc}$) for $\mathcal{E}_k = 10^{47.5} {\rm~erg}$ for ALMA (ngVLA), respectively.

In the right panels, we show the results for $f_e = 0.1$, $\epsilon_e = 0.1$, and $\epsilon_B = 0.1$, in which the detection horizon becomes larger. For example, in Fig.~\ref{fig:2Dplot_100GHz_30yr_0.1}, the values of $d_{L, \rm lim}$ for $T_{\rm age} = 30 {\rm~yr}$ are $0.026 {\rm~Mpc}$ ($0.13 {\rm~Mpc}$) for $\mathcal{E}_k = 10^{42} {\rm~erg}$, $3.8 {\rm~Mpc}$ ($20 {\rm~Mpc}$) for $\mathcal{E}_k = 10^{45} {\rm~erg}$ and $190 {\rm~Mpc}$ ($980 {\rm~Mpc}$) for $\mathcal{E}_k = 10^{47.5} {\rm~erg}$ for ALMA (ngVLA), respectively.

In Fig.~\ref{fig:2Dplot_1GHz}, we show similar plots for $\nu = 1\rm~GHz$. 
Because the observed flux at 1~GHz is more strongly affected by the effect of FF absorption than that at 100~GHz for $T_{\rm age} = 30 {\rm~yr}$, we only show the results for $T_{\rm age} = 100 {\rm~yr}$.
The detection horizons at the 1 GHz band for VLA (solid curve) and SKA (dashed curve) are indicated as black curves in the $d_{L}-\mathcal{E}_{k}$ plane.
Contrary to the 100 GHz band, the detection horizon at the 1 GHz band is comparable for bursts with a kinetic energy of $\mathcal{E}_k \gtrsim 10^{45} {\rm~erg}$ for both parameter sets of the microphysical parameters.

\subsection{Expected numbers of afterglows detected with current and next-generation radio telescopes}
\label{subsect:number}

The total number of detectable events, $N_{\rm burst}$, could be obtained from the following equation
\begin{equation}\label{eq:N}
N_{\rm burst} =  \Delta T  \int_{\mathcal{E}_{k,\rm min}}^{\mathcal{E}_{k,\rm max}} d \mathcal{E}_{k} \int_{0}^{d_{L, \rm lim}( \mathcal{E}_{k} )} dd_{L} 4 \pi d_{L}^2 \frac{dR_{\rm burst}}{d\mathcal{E}_{k}},
\end{equation}
where $\Delta T$ is the observation time of the telescope, 
$\mathcal{E}_{k, \rm min}$ is the minimum kinetic energy, 
$\mathcal{E}_{k, \rm max}$ is the maximum kinetic energy, 
$d_{L, \rm lim}( \mathcal{E}_{k} )$ is the detection horizon for a given $\mathcal{E}_{k}$ and $dR_{\rm burst} / d\mathcal{E}_{k}$ is the differential event rate of the bursts. 

\subsubsection{MGF scenario}
\label{sect:MGF}
In this section, we consider bursts driven by MGFs. 
According to~\cite{Burns_2021ApJ...907L..28B}, the best-fit volumetric rate of MGFs is $R_{\rm MGF} \approx 3.8 \times 10^{5} \ {\rm~Gpc^{-3}~yr^{-1}}$ assuming the energy distribution of MGFs follows a power-law distribution with spectral index $\alpha \approx 1.7$.
The lowest considered value of the isotropic-equivalent energy of the MGFs is $\mathcal{E}_{\rm iso, min} = 3.7 \times 10^{44} {\rm~erg}$ and the maximum isotropic-equivalent energy could be as high as $\mathcal{E}_{\rm iso, max} = 5.75 \times 10^{47} {\rm~erg}$~\citep{Burns_2021ApJ...907L..28B}.
Here, we assume $\mathcal{E}_{k} \approx \mathcal{E}_{\rm iso}$. 
This is conservative in the sense that the radiative efficiency is less than unity, e.g., $\mathcal{E}_{k} \sim 5\mathcal{E}_{\rm iso}$ for GRBs~\citep[e.g.,][]{Kumar:2014upa}.
For a given age $T_{\rm age}$, the burst rate density $dR_{\rm burst}/d\mathcal{E}_{k}$ can be estimated as
\begin{equation}\label{eq:gEk_MGF}
    \frac{dR_{\rm burst}|_{T_{\rm age}}}{d\mathcal{E}_{k}} = f_{\rm age} R_{\rm MGF} \frac{(1 - \alpha) \mathcal{E}_{k}^{-\alpha}}{\mathcal{E}_{\rm iso, max}^{1-\alpha} - \mathcal{E}_{\rm iso, min}^{1-\alpha}},
\end{equation}
where $f_{\rm age}=T_{\rm age}/T_{\rm life}$ represents the fraction of magnetars with $T_{\rm age}$ and $T_{\rm life} \sim 10^4 {~\rm yr}$ is the typical life time of magnetars~\citep[e.g.,][]{Kaspi_2017ARA&A..55..261K}. 
We assume that the value of $f_{\rm age}$ is 0.01 for $T_{\rm age} = 100 {\rm~yr}$ while it is 0.003 for $T_{\rm age} = 30 {\rm~yr}$.

\begin{table}
    \centering
    \begin{tabular}{c|c|c|c}
    \hline
    \multicolumn{2}{ c| }{\diagbox{$\nu$}{$T_{\rm age}$}} & 30 yr & 100 yr \\
    \hline
    1 GHz & VLA & 0 & 0.020  \\ 
    	 & SKA & 0 & 0.17 \\ 
    100 GHz & ALMA & 0.054 & 0.0021 \\ 
    	 & ngVLA & 7.6 & 0.35 \\ 
    	 \hline
    \multicolumn{2}{ c| }{\diagbox{$\nu$}{$T_{\rm age}$}} & 30 yr & 100 yr \\
    \hline
    1 GHz & VLA & 0 & 0.013  \\ 
    	 & SKA & 0 & 0.13 \\ 
    100 GHz & ALMA & 2.0 & 0.086 \\ 
    	 & ngVLA & 270 & 12 \\ \hline
    \end{tabular}
\caption{Expectation numbers of detectable afterglow events. The observational time window of radio telescopes is assumed to be $\Delta T = 10 \rm~yr$. 
In the upper and lower parts, we assume $f_e=0.05$, $\epsilon_e=0.01$, and $\epsilon_B = 0.001$, and $f_e=0.1$, $\epsilon_e = 0.1$, and $\epsilon_B = 0.1$, respectively.}
\label{table:DetectionNumber_N_MGF}
\end{table}

\begin{table}
    \centering
    \begin{tabular}{c|c|c|c}
    \hline
    \multicolumn{2}{ c| }{\diagbox{$\nu$}{$T_{\rm age}$}} & 30 yr & 100 yr \\
    \hline
    1 GHz & VLA & 0 & 0.020 \\ 
    	 & SKA & 0 & 0.17 \\ 
    100 GHz & ALMA & 0.0085 & 0.0021 \\ 
    	 & ngVLA & 1.4 & 0.35 \\ \hline
    \multicolumn{2}{ c| }{\diagbox{$\nu$}{$T_{\rm age}$}} & 30 yr & 100 yr \\
    \hline
    1 GHz & VLA & 0 & 0.013 \\ 
    	 & SKA & 0 & 0.13 \\ 
    100 GHz & ALMA & 0.21 & 0.085 \\ 
    	 & ngVLA & 36 & 12 \\ \hline
    \end{tabular}
\caption{Similar to Table~\ref{table:DetectionNumber_N_MGF}, but the difference is that here we assume that the observation is conducted 1 day after the burst.}
\label{table:DetectionNumber_N_MGF_1day}
\end{table}

In Table~\ref{table:DetectionNumber_N_MGF}, we show the total number of detected events using Eq.~\ref{eq:N} and Eq.~\ref{eq:gEk_MGF} with $\Delta T = 10 \rm~yr$.
We can see that the expected detection number is more prominent at the 100 GHz band by ALMA and ngVLA. 
In particular, ngVLA could detect $N_{\rm burst} \sim 7.6$ for $T_{\rm age} = 30 {\rm~yr}$ and $N_{\rm burst} \sim 0.35$ for $T_{\rm age} = 100 {\rm~yr}$ in the conservative case, while the corresponding number is $\sim 35$ times larger for $f_e = 0.1$, $\epsilon_e = 0.1$, and $\epsilon_B = 0.1$.
The detection number with ALMA is expected to be $N_{\rm burst} \sim 2$ for $f_e = 0.1$, $\epsilon_e = 0.1$, and $\epsilon_B = 0.1$ with $T_{\rm age} = 30 ~\rm yr$.
At the 1 GHz band, the expected detection number is less than unity for both VLA and SKA with $\Delta T = 10 {\rm~yr}$.
The results are consistent with the conclusions derived in the previous section where the detection at the 1 GHz band for smaller $T_{\rm age}$ is largely affected by the effect of FF absorption.
Note the effect of FF absorption could be reduced considering the spatial distribution of the SN ejecta is asymmetric or clumpy, where a fraction of photons is expected to escape freely.

Because the observation time of radio telescopes cannot be easily adjusted to the peak time of afterglow emission, we calculate the total number of detected events assuming that the observation is conducted $1\rm~day$ after the burst, for which the results are shown in Table~\ref{table:DetectionNumber_N_MGF_1day}.
We find that the corresponding values of $N_{\rm burst}$ are decreased by a factor $\sim 1 - 10$, but the expected detection number for ngVLA at the 100 GHz band could be $N_{\rm burst} \gtrsim 1$. 

\subsubsection{FRB scenario}

In this section, we consider the scenario, in which bursts are associated with FRBs. The burst rate density $dR_{\rm burst} / d\mathcal{E}_{k}$ can be estimated as
\begin{equation}\label{eq:gEk_FRB}
    \frac{dR_{\rm burst}|_{T_{\rm age}}}{d\mathcal{E}_{k}} = f_{\rm age}R_{\rm FRB} \frac{\mathcal{E}_k^{-\alpha}}{\mathcal{E}_{k, \rm cr}^{1 - \alpha}},
\end{equation}
where $R_{\rm FRB} \approx 339 {\rm~Gpc^{-3}~yr^{-1}}$ is the characteristic volumetric rate of FRBs at $\mathcal{E}_{k, \rm cr}$, where $\mathcal{E}_{k, \rm cr}$ is the critical kinetic energy and $\alpha \approx 1.79$ is the index~\citep{2020MNRAS.494..665L}. 
The critical energy can be estimated as $\mathcal{E}_{k, \rm cr} = (\delta t / \eta_{\rm FRB}) L_{\rm cr} \sim 2.9 \times 10^{46} \rm~erg$ with $\delta t / \eta_{\rm FRB} = 100\rm~s$, where $\delta t$ is the duration of FRB, $\eta_{\rm FRB}$ is the radiative efficiency and ${\rm log}_{10} (L_{\rm cr}[\rm~erg~s^{-1}]) \simeq 44.46$ is the logrithmic of the critical luminosity of GRB~\citep{2020MNRAS.494..665L}.
Here we use $\mathcal{E}_{k, \rm min} = 10^{42} {\rm~erg}$ and $\mathcal{E}_{k, \rm max} = 10^{47.5} {\rm~erg}$ to calculate $N_{\rm burst}$.

\begin{table}
    \centering
    \begin{tabular}{c|c|c|c}
    \hline
    \multicolumn{2}{ c| }{\diagbox{$\nu$}{$T_{\rm age}$}} & 30 yr & 100 yr \\
    \hline
    1 GHz & VLA & 0 & $4.8 \times 10^{-4}$ \\ 
    	 & SKA & 0 & 0.0041 \\ 
    100 GHz & ALMA & 0.0012 & $5.0 \times 10^{-5}$ \\ 
    	 & ngVLA & 0.18 & 0.0081 \\ \hline
    \multicolumn{2}{ c| }{\diagbox{$\nu$}{$T_{\rm age}$}} & 30 yr & 100 yr \\
    \hline
    1 GHz & VLA & 0 & $3.2 \times 10^{-4}$ \\ 
    	 & SKA & 0 & 0.0032 \\ 
    100 GHz & ALMA & 0.046 & 0.0020 \\ 
    	 & ngVLA & 6.3 & 0.28 \\ \hline
    \end{tabular}
\caption{Similar to Table~\ref{table:DetectionNumber_N_MGF},  but the detection number $N_{\rm burst}$ is obtained by assuming that the burst is driven by an FRB with $\Delta T = 10 \rm~yr$.}
\label{table:DetectionNumber_N_FRB_100}
\end{table}

\begin{table}
    \centering
    \begin{tabular}{c|c|c|c}
    \hline
    \multicolumn{2}{ c| }{\diagbox{$\nu$}{$T_{\rm age}$}} & 30 yr & 100 yr \\
    \hline
    1 GHz & VLA & 0 & $4.8 \times 10^{-4}$ \\ 
    	 & SKA & 0 & 0.0041 \\ 
    100 GHz & ALMA & $2.0 \times 10^{-4}$ & $4.9 \times 10^{-5}$ \\ 
    	 & ngVLA & 0.033 & 0.0080 \\ \hline
    \multicolumn{2}{ c| }{\diagbox{$\nu$}{$T_{\rm age}$}} & 30 yr & 100 yr \\
    \hline
    1 GHz & VLA & 0 & $3.2 \times 10^{-4}$ \\ 
    	 & SKA & 0 & 0.0032 \\ 
    100 GHz & ALMA & 0.0049 & 0.0020 \\ 
    	 & ngVLA & 0.82 & 0.27 \\ \hline
    \end{tabular}
\caption{Similar to Table~\ref{table:DetectionNumber_N_FRB_100}, but the difference is that here we assume that the observation is conducted 1 day after the burst.}
\label{table:DetectionNumber_N_FRB_100_1day}
\end{table}

The results are summarized in Table~\ref{table:DetectionNumber_N_FRB_100}-\ref{table:DetectionNumber_N_FRB_100_1day}.
Compared to Table~\ref{table:DetectionNumber_N_MGF}-\ref{table:DetectionNumber_N_MGF_1day}, the total number of detectable afterglow events decreases by a factor of $\sim 50$, which is consistent with the difference in the measured volumetric rate between MGFs and FRBs adopted in this work.

\section{Summary and Discussion}
In this work, we performed detailed studies on the afterglow emission caused by bursts that may be related to MGFs or FRBs occurring in their wind nebulae and surrounding baryonic ejecta.
Based on the bursts-in-bubble model, we used both analytical and numerical methods to calculate the dynamical evolution of the trans-relativistic shell formed when the nebula swept by the burst-driven outflow propagates within the dense baryonic ejecta, and the energy spectra and light curves of the afterglow emission from refreshed forward shock.
We adopted the thick-shell approximation considering the merged shell has an initial width $\Delta R_{\rm ej}$ after compressed by the FRB and/or MGF bursts.

Motivated by the detection of FRB 20200120E, we calculated the light curves and spectra assuming there is a burst occurred at $d_L=4$~Mpc with different kinetic-energy $\mathcal{E}_{k}$ and age $T_{\rm age}$. 
The detection of the afterglow emission for energetic bursts with $\mathcal{E}_{k} = 10^{47}{\rm~erg}$ is promising, where the flux is above the detection threshold of current and next-generation radio telescopes for reasonable values of $T_{\rm age}$. 
Due to free-free absorption, the detection of the afterglow emission with VLA and SKA at the 1 GHz band would be challenging when $T_{\rm age} \lesssim 30 {\rm~yr}$.

We also studied the detectability of afterglow with current X-ray telescopes, such as \textit{Chandra} and \textit{XMM-Newton}. 
Our results show that the X-ray telescopes have the ability to detect X-ray afterglow emission with $T_{\rm age} = 30 {\rm~yr}$ and $T_{\rm age} = 100 {\rm~yr}$ at $t \lesssim 1 {\rm~month}$ for $\mathcal{E}_{k} = 10^{47}~{\rm~erg}$. The detection of the afterglow with $T_{\rm age} = 30 {\rm~yr}$ ($100 {\rm~yr}$) at the optical band with the Vera C. Rubin Observatory is also promising at $t \lesssim 1 {\rm~month}$ ($1 {\rm~d}$) for $\mathcal{E}_{k} = 10^{47}~{\rm~erg}$. The joint observations at the multi-wavelength bands, including radio, optical, and X-ray, will allow us to unveil the details of the bursts-in-bubble model.

We then studied the detection horizon and expected number of events with current and next-generation radio telescopes.
We found the detection horizon $d_{L,\rm~lim}$ could range from 0.010 Mpc (0.059 Mpc) to 59 Mpc (310 Mpc) for bursts with kinetic energy from $10^{42} {\rm~erg}$ to $10^{47.5} {\rm~erg}$ for ALMA (ngVLA).
The detection horizon could be larger for microphysical parameters $f_e = 0.1$, $\epsilon_e = 0.1$, and $\epsilon_B = 0.1$.
The detection horizon at 1 GHz band with VLA and SKA is also promising for energetic bursts, except for the effect of FF absorption when $T_{\rm age} \lesssim 30\rm~yr$.

We calculated the number of detectable afterglow events for current and next-generation radio telescopes, as shown in Sec.~\ref{subsect:number}. 
If we assume that the burst rate follows MGFs, the expected numbers of burst afterglows for ALMA and ngVLA could be larger than unity with a few decades of operation. 
In the FRB scenario, the expected numbers would be less, but ngVLA has the possibility to detect the afterglow emission within a few ten years. 
However, radio telescopes such as ngVLA and ALMA cannot follow up on a burst immediately, so we also consider the observation is conducted $1\rm~day$ after the burst, as shown in Table~\ref{table:DetectionNumber_N_MGF_1day} and Table~\ref{table:DetectionNumber_N_FRB_100_1day}. We found the detection number becomes smaller than the ideal case, but still possible for ngVLA to detect the afterglow driven by MGFs and/or FRBs with suitable parameters within a few ten years. Our theoretical expectation regarding the number of afterglows detected with ALMA and VLA is consistent with the current observation that no similar signals are detected by these two radio telescopes. Note that our results on the expected numbers of afterglow depend on $T_{\rm age}$, and here we just show the results with $T_{\rm age} = 30 {\rm~yr}$ and $T_{\rm age} = 100 {\rm~yr}$. The values of $\mathcal{E}_{k, \rm min}$ and $\mathcal{E}_{k, \rm max}$ we adopted in this work for both MGFs and FRBs could affect the number of detected events, where future observations are needed to unveil the mysteries.

In this study, we focused on the forward shock component in afterglow emission.  In addition, the reverse shock component could be significant in the thick-shell limit around $t_\times\approx t_{\rm cross}$, which could enhance the flux at the radio band. The afterglow emission may also be outshined by the quasi-steady nebular emission. According to \cite{Murase:2016sqo}, if the nebula is powered by magnetar rotation, the observed energy flux of the nebular emission $\nu F^{\rm nb}_\nu$ depends on the spin-down power of the magnetar $L_{\rm sd} \simeq 2.4 \times 10^{39} B_{*,15}^2 P_{-0.5}^{-4} {\rm~erg~s^{-1}}$, where $B_*$ is the magnetic filed of magnetar, $P = P_i (1 + T_{\rm age}/T_{\rm sd})^{1/2}$ is the rotation period of magnetar, $P_i$ is the initial rotation period and $T_{\rm sd} \simeq 2.5 {~\rm yr} \ B_{*,15}^{-2} P_{i, -0.5}^2$ is the spin-down time. We found with $P_i = 1-10 \rm~s$, the nebular emission can be lower than afterglow fluxes for energetic bursts with $\mathcal{E}_k \gtrsim 10^{44} - 10^{45} ~\rm erg$ given that $f_e = 0.1$, $\epsilon_e = 0.1$, and $\epsilon_B = 0.1$. 
The nebular emission is model dependent, and the number of detectable afterglow events would be dominated by sufficiently energetic bursts, in which the nebular component is subdominant.

\section*{Acknowledgements}
We acknowledge the anonymous referee for valuable comments
The work was partly supported by the NSF grants No.~AST-1908689, No.~AST-2108466 and No.~AST-2108467, and KAKENHI No.~20H01901 and No.~20H05852 (K.M.).

\section*{Data Availability}
The data developed for the calculation in this work is available upon request.



\bibliographystyle{mnras}
\bibliography{kmurase2}




\bsp	
\label{lastpage}
\end{CJK*}
\end{document}